\documentclass[%
 reprint,
superscriptaddress,
nofootinbib,
 amsmath,amssymb,
 aps,
prd,
floatfix,
]{revtex4-2}

\usepackage{subcaption,epic}
\usepackage{tikz}

\usepackage{graphicx}
\usepackage{dcolumn}
\usepackage{bm}
\usepackage{xcolor}
\usepackage{comment}
\usepackage{lipsum}
\usepackage{multirow}
\usepackage{soul}
\usepackage{orcidlink}

\usepackage[newcommands]{ragged2e}

\usepackage[labelformat=simple, margin=1.0mm]{subcaption}
\DeclareCaptionJustification{justified}{\justifying}

\usepackage{float}
\usepackage{hyperref}
\hypersetup{
    colorlinks=true,
    linkcolor=blue,
    filecolor=magenta,      
    urlcolor=cyan,
    pdftitle={Methods},
    pdfpagemode=FullScreen,
    }

\newcommand{\rvline}{\hspace*{-\arraycolsep}\vline\hspace*{-\arraycolsep}}

\usepackage[colorinlistoftodos]{todonotes}

\begin{document}

\preprint{APS/123-QED}

\title{Improving Neutrino-Nuclei Interaction Models: Recommendations and Case Studies on Peelle’s Pertinent Puzzle}


\date{\today}

\newcommand{\Bern}{University of Bern, CH-3012 Bern, Switzerland}
\newcommand{\Brookhaven}{Brookhaven National Laboratory, Upton, NY 11973, USA}
\newcommand{\Columbia}{Columbia University, New York, NY 10027, USA}
\newcommand{\Fermi}{Fermi National Accelerator Laboratory, Batavia, IL 60510, USA}
\newcommand{\Imperial}{Imperial College London, Department of Physics, London SW7 2BZ, United Kingdom}
\newcommand{\LosAlmos}{Los Alamos National Laboratory, Los Alamos, NM 87545, USA}
\newcommand{\Michigan}{University of Michigan, Ann Arbor, MI 48109, USA}
\newcommand{\MichiganState}{Michigan State University, East Lansing, MI 48824, USA}
\newcommand{\Minntwin}{University of Minnesota Twin Cities, Minneapolis, MN 55455, USA}
\newcommand{\Oxford}{University of Oxford, Oxford, OX1 3RH, United Kingdom}
\newcommand{\Pitt}{University of Pittsburgh, Pittsburgh, PA 15260, USA}
\newcommand{\Rochester}{University of Rochester, Rochester, NY 14627, USA}
\newcommand{\Royalholloway}{Royal Holloway University of London, Egham, TW20 0EX, United Kingdom}
\newcommand{\Rutherford}{STFC Rutherford Appleton Laboratory, Didcot OX11 0QX, United Kingdom}
\newcommand{\TelAviv}{Tel Aviv University, Tel Aviv-Yafo, Israel}
\newcommand{\Texasaustin}{University of Texas at Austin, Austin, TX 78712, USA}\newcommand{\Tokyo}{Kamioka Observatory, Institute for Cosmic Ray Research, University of Tokyo, Kamioka, Gifu 506-1205, Japan}

\affiliation{\Bern}
\affiliation{\Brookhaven}
\affiliation{\Columbia}
\affiliation{\Fermi}
\affiliation{\Imperial}
\affiliation{\LosAlmos}
\affiliation{\Michigan}
\affiliation{\MichiganState}
\affiliation{\Minntwin}
\affiliation{\Oxford}
\affiliation{\Pitt}
\affiliation{\Rochester}
\affiliation{\Royalholloway}
\affiliation{\Rutherford}
\affiliation{\TelAviv}
\affiliation{\Texasaustin}
\affiliation{\Tokyo}

\author{S.~Abe} \affiliation{\Tokyo}
\author{L.~Aliaga-Soplin} \affiliation{\Texasaustin}
\author{J.~Barrow} \affiliation{\Minntwin}
\author{L.~Bathe-Peters} \affiliation{\Oxford}
\author{B.~Bogart} \affiliation{\Michigan}
\author{L.~Cooper-Troendle}\email[Corresponding author: ]{lcoopert@proton.me} \affiliation{\Pitt}
\author{R.~Diurba} \affiliation{\Bern}
\author{S.~Dytman} \affiliation{\Pitt}
\author{S.~Gardiner} \affiliation{\Fermi}
\author{L.~Hagaman} \affiliation{\Columbia}
\author{M.~S.~Ismail} \affiliation{\Pitt}
\author{J.~Isaacson} \affiliation{\Fermi}\affiliation{\MichiganState}
\author{J.~Kim} \affiliation{\Rochester}
\author{L.~Liu} \affiliation{\Fermi}
\author{J.~McKean} \affiliation{\Imperial}
\author{N.~Nayak} \affiliation{\Brookhaven}
\author{A.~Papadopoulou} \affiliation{\LosAlmos}
\author{L.~Pickering} \affiliation{\Rutherford}
\author{X.~Qian} \affiliation{\Brookhaven}
\author{K.~Skwarczy\'{n}ski} \affiliation{\Royalholloway}
\author{J.~Tena Vidal} \affiliation{\TelAviv}
\author{J.~Wolfs} \affiliation{\Rochester}

\begin{abstract}
Improving the modeling of neutrino-nuclei interactions using data-driven methods is crucial for high-precision neutrino oscillation experiments. This paper investigates Peelle’s Pertinent Puzzle (PPP) in the context of neutrino measurements, a longstanding challenge to fitting theoretical models to experimental data. Inconsistencies in data-model comparisons hinder efforts to enhance the accuracy and reliability of model predictions. We analyze various sources contributing to these inconsistencies and propose strategies to address them, supported by practical case studies. We advocate for incorporating model fitting exercises as a standard practice in cross section publications to enhance the robustness of results. We use a common analysis framework to explore PPP-related challenges with \texttt{MicroBooNE} and \texttt{T2K} data in an unified manner. Our findings offer valuable insights for improving the accuracy and reliability of neutrino-nuclei interaction models, particularly by systematically tuning models using data.
\end{abstract}

\maketitle

\section{Introduction}\label{sec:intro}

Current and next-generation accelerator neutrino oscillation experiments~\cite{T2K:2023smv,NOvA:2023iam,DUNE:2020jqi,Hyper-Kamiokande:2018ofw} in the GeV neutrino energy range hold the potential to address some of the most pressing questions in neutrino physics, such as the matter-antimatter asymmetry through charge-parity violation~\cite{Abe:2019vii}, the precise ordering of neutrino masses among the three generations~\cite{Qian:2015waa}, and the search for sterile neutrinos~\cite{Abazajian:2012ys}. These experiments, particularly DUNE~\cite{DUNE:2020jqi} and Hyper-Kamiokande~\cite{Hyper-Kamiokande:2018ofw}, aim to provide unprecedented precision in measuring neutrino oscillation parameters. However, achieving such precision necessitates a comprehensive understanding and meticulous control of the systematic uncertainties affecting these measurements~\cite{DiLodovico:2023jgr}.

One of the critical sources of systematic uncertainties in these experiments is the modeling of interaction between neutrinos and nuclei~\cite{Workman:2022ynf, Formaggio:2012cpf, Mosel:2016cwa}.
Neutrino-nuclei interactions are inherently complex because, while the interaction itself is governed by the electroweak force, the nuclear structure of the target is shaped by the strong force and therefore requires an understanding of the non-perturbative nature of quantum chromodynamics (QCD) in this energy regime. This complexity leads to incomplete theoretical descriptions and necessitates the use of effective models to describe these interactions~\cite{Gross:2022hyw}.

To mitigate these uncertainties, it is common practice to tune cross section models through the use of event generators. These event generators, such as \texttt{GENIE}~\cite{GENIE:2021npt}, \texttt{NEUT}~\cite{Hayato:2021heg}, \texttt{NuWro}~\cite{Golan:2012rfa}, \texttt{GiBUU}~\cite{Buss:2011mx}, and \texttt{ACHILLES}~\cite{PhysRevD.107.033007}, incorporate various interaction models to describe neutrino-nuclei interactions. These models are built under different assumptions of the nuclear ground state, its structure function as well as the dynamics of intra-nuclear absorption and scattering of various particles produced in the tree level interaction. Tuning these models based on experimental data allows researchers to align theoretical predictions, which often include necessary approximations, with real observations, thereby improving the reliability of predictions. Examples of such tuning efforts include the work done by \texttt{MINOS}~\cite{GALLAGHER2006229}, \texttt{T2K}~\cite{T2K:2023smv}, \texttt{NOvA}~\cite{NOvA:2020rbg}, \texttt{MINERvA}~\cite{MINERvA:2019kfr}, and \texttt{MicroBooNE}~\cite{MicroBooNE:2021ccs}. This process not only enhances the precision of neutrino oscillation experiments but also advances our understanding of neutrino scattering and nuclear physics.

In this paper, we investigate Peelle’s Pertinent Puzzle~\cite{fruhwirth2012peelle} (PPP), a phenomenon where unexpected normalization values, often smaller than anticipated, are obtained from fitting a model to experimental data with correlated systematic uncertainties. An example of this is the reactor antineutrino anomaly~\cite{Mention:2011rk, Zhang:2023zif}, where the original best-fit data-to-prediction ratio is lower than the intuitive mean, as highlighted in Ref.~\cite{Zhang:2013ela}. In this specific case, it is possible to reformulate the fit to avoid a PPP issue~\cite{GIUNTI2022137054}, but this procedure does not generalize well to all situations. PPP-related behavior has also been observed in the model fitting of neutrino-nuclei cross section data~\cite{Chakrani:2023htw}. While traditionally the study of PPP has focused on erroneous deviations in the normalization of a model fit to data, we use a broader definition here to more comprehensively address inaccurate fits to data. In this work, we use PPP to refer to any fit degradation that results from an improper treatment in the data with its full covariance, model, or their comparison. This definition includes fit distortions outside of the overall normalization, which we refer to as ``non-normalization PPP".

As discussed in Ref.~\cite{fruhwirth2012peelle}, improper estimates of uncertainties, such as covariance matrices, can contribute to PPP. Generally, PPP arises from inconsistencies between experimental data and theoretical models, and can serve as an indicator of such inconsistencies. However, the absence of PPP does not necessarily indicate consistency between experimental data and theoretical models. Other metrics beyond the normalization of the fitted mean, such as the $p$-value, can also serve as indicators of an inconsistency.

In this work, we examine PPP through case studies of model fitting of neutrino-nuclei cross section data and propose recommendations to enhance the fidelity of tuning neutrino-nuclei interaction models with experimental data. In Sec.~\ref{sec:mismatch}, we discuss how inconsistencies between data and models can introduce PPP and provide recommendations on how to avoid such inconsistencies. In Sec.~\ref{sec:cov_matrix}, we describe the covariance matrix formalism approach to fitting models to cross section data, highlighting its flexibility. In Sec.~\ref{sec:methodology} we describe methods that can help detect cases of PPP. In Sec.~\ref{sec:mitigation} we describe various mitigation strategies that can enable successful model fitting in the presence of PPP issues. In Sec.~\ref{sec:studies}, we detail case studies of fitting neutrino-nuclei interaction models to experimental data and discuss how to identify data-model mismatches beyond abnormal best-fit normalization. Finally, in Sec.~\ref{sec:recommend}, we present strategies to address PPP when it is detected, and we summarize our findings in Sec.~\ref{sec:summary}.

\section{Inconsistencies between data and models}\label{sec:mismatch}

Accurate comparisons between experimental data and theoretical models are fundamental to improve our understanding of neutrino-nucleus interactions. However, mismatches often arise due to limitations in the models, biases in the data, or inconsistencies in how the two are defined, interpreted, or communicated. These discrepancies can obscure underlying physics, distort model tuning efforts, and ultimately hinder progress in understanding fundamental interactions. This section examines the origins and implications of data-model inconsistencies, beginning with an illustrative case of parameter pull pathology, before exploring model limitations, the validity of Gaussian approximations, and communication mismatches in the interpretation and reporting of experimental results. Together, these discussions highlight both conceptual and practical challenges that must be addressed to ensure robust and reliable comparisons between theory and data.

\begin{figure}[!htbp]
    \centering
    \includegraphics[width=1.0\linewidth]{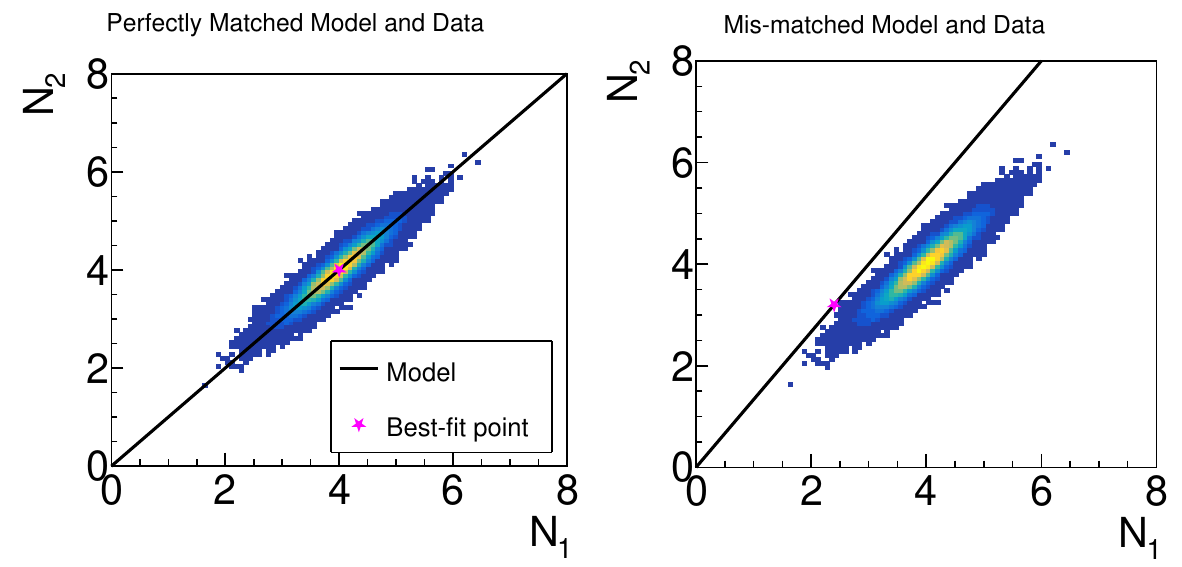}
    \caption{ Illustration of PPP with a simple two-dimensional measurement. Data preference is represented by the color contour, model preference is represented by the black line, and the model best-fit point is represented by the magenta star. Scenarios of data-model agreement and data-model mismatch are shown on the left and right, respectively. }
    \label{fig:ppp_illustration}
\end{figure}

The PPP phenomenon, which is often visually apparent from an abnormally low best-fit normalization, may occur under the following conditions: (i) strong correlations within the measurements of a dataset, (ii) model limitations, and (iii) an inconsistency between model predictions and data preferences. Often, a data-model inconsistency, also referred to as a mismatch, tension, or incompatibility, will lead to a high $\chi^{2}$ value in the goodness-of-fit (GoF) assessment for the best-fit model. This allows high-$\chi^{2}$ GoF to be taken as a symptom indicative of a potential data-model inconsistency and a PPP issue, although causation cannot be inferred with certainty. However, even in the absence of any visible PPP issues, a high $\chi^{2}$ value still reflects significant data-model disagreement, posing a major challenge for effective model tuning. In Fig.~\ref{fig:ppp_illustration}, we illustrate the relationship between PPP and inconsistencies between data and the model through fitting a model with only one degree of freedom, the normalization of $N_{1}+N_{2}$, to a two-dimensional measurement. The right panel shows a case of a data-model inconsistency, where the model is unable to describe the relation between $N_{1}$ and $N_{2}$ in the data, resulting in a poor overall normalization in the fitted model after minimizing the $\chi^{2}$.

Several scenarios can lead to mismatches between the data and the model. These may stem from deficiencies in the model, inaccuracies or biases in the data, or inconsistencies in how the model predictions and data are defined and interpreted. In the following sections, we will examine each of these possibilities in more detail.

\subsection{Model Limitations}
When addressing issues related to the model, we limit our discussion by excluding cases where the model is entirely unsuitable. Instead, we focus specifically on situations where the model has insufficient range of model parameters, or degrees of freedom, needed to fully capture the underlying physics. 
Modern event generators used to model neutrino-nuclei interactions typically include dozens of parameters that control the simulation of neutrino-nucleus interactions, but still rely on approximations in their modeling. Sometimes, parameter uncertainty estimates are provided by event generator experts, helping to compensate for model limitations. Additionally, it is possible for computational limitations to restrict the ability to fully explore the parameter space available in a model during fitting. When many parameters must be examined to find the minimum $\chi^{2}$ test statistic for a best-fit search, the accumulated time required to generate new predictions for each set of parameters in a high-dimensional space can become impractical. One strategy in such cases is to focus on a selection of parameters deemed most relevant, while keeping all other parameters' central values fixed at their nominal values. This approach can often be effective, but it is not universally applicable. Fixing certain parameters' central values at their nominal values restricts the overall phase space and can lead to sizable mismatches between the data and the model, particularly when non-negligible parameters are fixed. In Sec.~\ref{sec:cov_matrix}, we present a covariance matrix-based fitting procedure that enables the simultaneous inclusion of numerous parameters without substantially increasing computational time.

\subsection{Improper Treatments of Data}
In addition to addressing issues related to the model, it is equally important to consider potential errors and approximations in the measurement of the data itself. These data frequently rely on Gaussian approximations when reporting their results. For example, neutrino-nuclei interaction cross section models are used to help unfold from reconstructed quantities to a measurement over truth quantities. These models are typically presented as central values accompanied by a multivariate Gaussian distribution defined by a covariance matrix. While the validity of this approximation remains a subject of active discussion, particularly for results driven by systematic uncertainties~\cite{Chakrani:2023htw,DAgostini:1993arp,radev2024flow}, model validation procedures such as those outlined in Refs.~\cite{MicroBooNE:2024kwe,MicroBooNE:2021sfa,MicroBooNE:2023foc,MicroBooNE:2024xod} provide effective means to mitigate these concerns for unfolding models used in producing data measurements. There are also other validation tools that can help form a comprehensive validation approach, such as the use of fake data studies to examine potential model dependence and side bands to examine the modeling of backgrounds.

Depending on the underlying Monte Carlo simulation model, the effects of systematic uncertainties may not always conform to a Gaussian approximation due to non-linear and asymmetric relations between model parameters and the overall model prediction. These behaviors deviate from the Gaussian approximation, which can introduce discrepancies in evaluating a \textit{p}-value when comparing results from the original base model to those derived using the covariance matrix.  However, as demonstrated in Ref.~\cite{radev2024flow}, this Gaussian approximation tends to be overly aggressive at estimating low \textit{p}-values, underestimating their true value. As a result, the Gaussian approach is more sensitive to model discrepancies in the low \textit{p}-value regime, meaning that a test under this approximation with a \textit{p}-value threshold such as 0.05 is conservative compared to one using the true \textit{p}-value. This means it is possible to accurately perform model validation tests under the Gaussian approximation to check whether the model can describe the data within its uncertainties. In this context, while the fidelity of the covariance-matrix approximation to the underlying model and the validity of the Gaussian approximation remain important, greater emphasis should be placed on its ability to robustly capture the critical features relevant to the analysis. Data-driven model validation, such as the methodologies presented in~\cite{MicroBooNE:2024kwe}, can be especially helpful in this task.

\subsection{Improper Data-Model Comparisons}
So far, in the context of PPP, we've discussed data-model mismatches arising separately from either the model or the unfolded measurement, both from the Gaussian approximation of its uncertainties or during the unfolding procedure itself. Beyond these, data-model mismatches can also arise from an incorrect comparison between the model prediction and the unfolded data result, based on the formalism used to derive either of them.
Two such examples include the regularization matrix, $A_C$, involved in the unfolding procedure~\cite{Tang:2017rob,gardiner2024} that accounts for the effects of regularization and the flux over which the cross-section result is averaged over, either the real unknown flux or the nominal central value flux~\cite{Koch:2020oyf,MicroBooNE:2024kwe}.

In the extraction of cross sections from experimental data, properly accounting for detector resolution and efficiency effects is essential. One common approach is the use of an unfolding procedure that corrects for the smearing of measured quantities. In this case, to prevent non-intuitive results arising from the non-invertibility of this smearing, regularization techniques, such as imposing smoothness or non-negativity constraints, are commonly applied. The influence of regularization on the final unfolding results is captured in the regularization matrix, originally defined in the context of Wiener SVD unfolding and expanded to other methods in Ref.~\cite{gardiner2024}. Publishing this matrix is critical, as it eliminates the need for introducing an ad hoc and often poorly defined ``unfolding error". Providing the $A_C$ matrix allows model tuning teams to apply it directly to their models, ensuring consistent and fair comparisons between data and model predictions. In contrast, the absence of the $A_C$ matrix can create discrepancies between data and model results, even if an unfolding regularization uncertainty is provided in the absence of the exact $A_C$ matrix. This can complicate tuning efforts and reduce the reliability and utility of the extracted cross sections. As an alternative, avoiding regularization altogether in the unfolding process requires using wider binning, which, while sidestepping regularization effects, often results in the loss of valuable information. Another approach is to report both regularized and unregularized measurements, such as in Ref.~\cite{PhysRevD.98.032003}, although this essentially passes the problem of choosing the unfolding approach and corresponding tradeoffs onto anyone who intends to use the measurement.

During model tuning, the model prediction relies on a known neutrino spectrum input, typically the nominal neutrino flux reported by experiments. However, the measured event counts in a given kinematic bin are observed under an unknown real neutrino spectrum to which the experiment is exposed. The difference between the real and nominal neutrino spectrum is incorporated into the reported neutrino flux uncertainties. Since extracted cross sections are derived from these measured counts, their central values inherently depend on the unknown real neutrino spectrum. To avoid mismatches between the data and the model, it is crucial to reconcile these differences. This can be achieved by extrapolating the measured cross sections to the nominal neutrino spectrum or vice versa. As shown in Refs.~\cite{MicroBooNE:2021sfa,MicroBooNE:2023foc,MicroBooNE:2024xod}, cross sections can be extracted directly at the nominal neutrino flux, provided that the mapping between the neutrino flux and the visible kinematic variables is supported by a rigorously validated extrapolation model. Alternatively, the correlation between the extracted cross sections and the neutrino flux model can be reported~\cite{MicroBooNE:2024kwe}, allowing model tuning teams to account for uncertainties when extrapolating from cross sections at the nominal spectrum to those at the true spectrum, while taking into account the correlation with reported cross section results. Failure to address these differences and correlations can lead to significant mismatches~\cite{MicroBooNE:2024kwe}, as comparing model predictions for cross sections at a nominal neutrino spectrum with measured cross sections based on an unknown real neutrino spectrum introduces inconsistencies and reduces the reliability of the analysis~\cite{Koch:2020oyf}.

\section{Covariance Matrix Formalism in Fitting Models to Cross Section Data}~\label{sec:cov_matrix}
This section describes the conditional covariance matrix formalism, which can be used to fit models to data. 
Given two sets of random variables (X, Y) and the full covariance matrix $\Sigma$ describing their correlations: 
\begin{equation}
\Sigma = \begin{pmatrix}
\Sigma^{XX} & \Sigma^{XY} \\
\Sigma^{YX} & \Sigma^{YY}
\end{pmatrix},
\end{equation}
where $n$ is used to describe the measurement vector and $\mu$ is used to describe the prediction vector. We can then derive the prediction for \( X \) given the constraints on \( Y \) provided by a measurement:
\begin{eqnarray}\label{eq:conditional_mean_var}
\mu^{X,\text{const.}} &=& \mu^{X} + \Sigma^{XY} \cdot (\Sigma^{YY})^{-1} \cdot (n^Y - \mu^Y) \nonumber \\
\Sigma^{XX,\text{const.}} &=& \Sigma^{XX} - \Sigma^{XY} \cdot (\Sigma^{YY})^{-1} \cdot \Sigma^{YX}.
\end{eqnarray}
This is a general result given that X and Y are jointly Gaussian distributed and can be obtained by conditioning one distribution by the other~\cite{gp_rasmussen}. When using the covariance matrix formalism in fitting models to cross section data, we define X with dimension  $m_X$  to represent the model parameters under tuning, and  Y  with dimension  $m_Y$  to represent the actual cross section prediction in a given binning. The full covariance matrix $\Sigma$ is then given by:
\begin{equation}
\Sigma = \Sigma_{n} + \Sigma_{\mu},
\end{equation}
where $\Sigma_{n}$ is obtained by expanding the original  $m_Y$-dimensional measured cross section covariance matrix to $m_{X}+m_{Y}$ dimensions by filling zeros in the additional  $m_X$  dimensions. The $\Sigma_{\mu}$ represents the model covariance matrix and is obtained by simulating different universes of cross section predictions while varying the model parameters. With a sufficient number of universes simulated and their model parameters recorded, the model covariance matrix, including both model parameters (X) and model predictions of cross sections (Y), is constructed following the standard definition of a covariance matrix. 

This covariance matrix formalism has both advantages and limitations. The primary advantage is computational efficiency. By generating different universes of model predictions in parallel and constructing the overall covariance matrix beforehand, one can directly calculate the best-fit solution, thus significantly reducing the time for the actual fitting process. Once the best-fit parameters are obtained, the conditional covariance matrix formalism can be reused to make a constrained model prediction based on these parameters. Since the computational effort is dominated by the number of universes rather than the number of model parameters, this formalism allows for easy marginalization and fitting of multiple model parameters with fewer concerns about hyperparameters and fit instability, thereby enhancing the robustness and scalability of the model-fitting process. This procedure naturally addresses the concern of insufficient model coverage discussed in Sec.~\autoref{sec:mismatch}.

However, this formalism also has a few limitations. In particular, the conditional mean and variance approach given by Eq.~\ref{eq:conditional_mean_var} does not naturally enforce physical boundaries on model parameters (e.g., constraints such as non-negativity or bounded ranges). One alternative is to construct a test statistic using the full joint covariance matrix $\Sigma$ over both X and Y:
\begin{equation}\label{eq:cov_test_statistics}
T(X) = \left(X-\mu^X, n^Y-\mu^Y \right) \cdot \Sigma^{-1} \cdot \left(X-\mu^X, n^Y-\mu^Y \right)^T,
\end{equation}
where $n^Y$ and $\mu^Y$ are the observed and expected values of $Y$, respectively; $X$ is the model parameter being varied during the minimization, while $\mu^X$ is its expected value under the nominal model and remains fixed. In this formulation, $\mu^Y$ is treated as fixed, consistent with the expectation under the nominal model. The test statistic $T(X)$ can be minimized with software such as Minuit2~\cite{James:2004xla}, allowing the inclusion of physical boundaries on $X$ via the minimizer’s configuration. Although the conditional approach is computationally efficient and suitable for many applications, using the full joint covariance matrix can be helpful in cases where enforcing such constraints is essential. Additionally, since the covariance matrix $\Sigma$ is independent of the minimization and can be inverted ahead of time, the evaluation of $T(X)$ remains computationally efficient. We recommend that analysts enforce physical boundaries on parameter values in this manner so that meaningful fits are produced.

Furthermore, as discussed in Sec.~\ref{sec:mismatch}, the use of a covariance matrix naturally assumes that the underlying distribution follows a multivariate Gaussian distribution, which may deviate from the original model. This implies that the minimum found by minimizing $T(X)$, or obtained through the conditional covariance-matrix formalism, may differ from the true minimum defined by comparing the original model prediction with the data observation using only the data covariance matrix. Generally, this discrepancy should not pose a problem as long as the two minima are reasonably close. Therefore, we advocate checking the difference in parameter values between these two minima. In practice, one can initially fix the less relevant parameters at their best-fit values and allow only the important parameters to vary during the original model fit. Once the minimum is found, the comparison between the two minima can be performed with respect to the derived uncertainties on the best-fit parameters from the conditional covariance-matrix model fit. An example of the covariance matrix model fitting approach was created to support this work. It uses one of the models and measurements considered in this paper and is provided on Github~\cite{conditional_constraint_example}.

\section{Detecting Tensions between Data and Models}\label{sec:methodology}

There are many reasons why a model prediction may be in tension with a measurement. If the tension arises from inaccuracies in the neutrino interaction modeling, such comparisons can provide valuable insights for refining these models. Neutrino interaction models, when constructed with sufficient uncertainties and flexibility, allow for meaningful tuning. Even if the \textit{a priori} central value prediction significantly differs from the measurement, the fitting process can yield a reasonable $\chi^{2}/\mathrm{ndf}$ comparison and provide useful updates to the model parameters. However, tensions may also result from improper treatments, such as those discussed in Sec.~\ref{sec:mismatch}. In these cases, the fitting process may fail, resulting in inaccurate or unphysical parameter values.

In practice, it is difficult to distinguish between cases where data-model tension arises from genuine inaccuracies in the neutrino interaction modeling and cases where it stems from improper treatments. While this distinction cannot be made with certainty without full knowledge of the methodologies used in the measurement and the model, a systematic approach, similar to the data-driven model validation procedure~\cite{MicroBooNE:2024kwe}, can help identify likely inconsistencies in cases where unexpected large discrepancies are encountered. This approach attempts to assess whether a data measurement can be accurately unfolded through the use of a supporting model which includes the Gaussian approximation of its uncertainties through a covariance matrix. This validation is achieved by testing whether the unfolding model can accurately describe the data within its stated uncertainties. The procedure also utilizes the conditional constraint formalism, similar to the one employed in Sec.~\ref{sec:cov_matrix}, which incorporates correlations between observables to refine model predictions and reduce uncertainties based on observed data in reconstructed space. This approach uses Bayes' theorem to update the model prediction on one observable given a constraint from a measurement on a different observable. The result is a more sensitive test of the model used to unfold the data into truth space. Note that this updated model prediction is only used in the validation test and not the unfolding procedure.

Similar to how data unfolding can benefit from an examination of data-model consistency, attempts to fit models to data can benefit from a detailed examination of data-model comparisons to assess their plausibility. We recommend employing a goodness-of-fit metric to detect potential mismatches due to improper treatments between a model and a measurement after fitting the model to the data. A straightforward method involves constructing a (global) $\chi^{2}$ test statistic using the model prediction, including its uncertainties, measurement, and corresponding covariance matrices, and calculating the associated \textit{p}-value. This \textit{p}-value quantifies the probability of observing a level of tension at least as extreme as the measured one if the model is correct. If the \textit{p}-value is sufficiently low, such as below 0.05 (corresponding to a significance above 2$\sigma$), it may indicate that the data-model comparison is invalid due to an inconsistency, which can be caused by an improper treatment in the analysis.

This global $\chi^2$ approach can sometimes detect improper treatments but may also be overly conservative in other cases. As illustrated in Fig.~\ref{fig:ppp_illustration}, a PPP-driven mismatch in a measurement-prediction comparison can result in a best fit with a significant difference in normalization. In such cases, the global $\chi^{2}/\mathrm{ndf}$ may fail to detect an issue, as poor agreement in a single degree of freedom (e.g., normalization) might not significantly impact the overall $\chi^{2}$. To address this, the combined error covariance matrix $\Sigma_{C}$ can be diagonalized through a basis change:
\begin{equation}
\Sigma_{C} = \Sigma_{n} + \Sigma_{\mu}, \quad \Sigma_{D} = R^{-1} \Sigma_{C} R,
\end{equation}
where $\Sigma_{n}$ and $\Sigma_{\mu}$ are the measurement and model covariance matrices, and $R$ is the matrix of eigenvectors of $\Sigma_{C}$. In this diagonalized basis, the difference between measurement and prediction becomes:
\begin{equation}
D = R^{-1}(n-\mu).
\end{equation}
Here, correlations between bins are removed, allowing the direct evaluation of individual bin contributions $\chi^{2}_{i}$ to the total $\chi^{2}$:
\begin{equation}
\chi^{2}_{i} = (D^{T} \Sigma_{D}^{-1} D)_{i} = D^{2}_{i} / \Sigma_{D,ii},
\end{equation}
where $D_i$ is the $i$th element of the vector $D$, and $\Sigma_{D,ii}$ is the $i$th diagonal component of the matrix $\Sigma_D$. Note that it is possible for degrees of freedom (DoF) other than normalization to exhibit tension between data and model, which may indicate a ``non-normalization PPP" issue. If only one or a few DoF contribute disproportionately large $\chi^{2}_{i}$ terms, the total $\chi^{2}/\mathrm{ndf}$ may remain below threshold, making the global comparison insensitive to specific mismatches. 

Building on this observation, it is possible to identify mismatches at a local level by examining the $\chi^{2}_{i}$ values and corresponding \textit{p}-values of individual bins. It is important to control the family-wise error rate by correcting local \textit{p}-values when comparing them to a significance threshold to account for the look-elsewhere effect~\cite{Gross:2010qma} and avoid over-reporting extreme values.  There are many useful correction procedures; here we apply the \v{S}id\'{a}k correction~\cite{Sidak:1967} to each local \textit{p}-value:
\begin{equation}
p_{\mathrm{corrected}} = 1 - (1-p_{\mathrm{local}})^{N},
\end{equation}
where $N$ represents the total number of local \textit{p}-values (bins). If any corrected \textit{p}-values fall below the stated threshold of 0.05, the measurement-prediction comparison can be rejected on the grounds of suspected improper treatment. This approach is particularly effective when mismatches are confined to a subset of the degrees of freedom in the measurement space. By focusing on individual bins in the diagonalized space, this additional examination is more likely to detect localized discrepancies that may be missed in a global comparison.

\section{Mitigating Inconsistencies between Data and Models}\label{sec:mitigation}
Once serious tension between data and model is detected such that an improper treatment is considered likely, the next step is to determine the appropriate course of action. One option is to abandon attempts to fit the model to the measurement, acknowledging that the inconsistencies render the process fundamentally flawed. Alternatively, if one believes that the model contains inaccuracies that are not sufficiently covered by parameter uncertainties, one can enlarge the model uncertainties. However, typically models are already assigned conservatively large uncertainties by default, so it may be difficult to sufficiently enlarge model uncertainties in a plausible manner. Finally, one can adopt a mitigation strategy aimed at addressing the presence of the tension, thereby enabling a meaningful and constructive comparison between the measurement and the model prediction. In this section we discuss potential mitigation strategies under the assumption that there is a PPP issue that must be addressed in a fit attempt as best as possible.

\subsection{Mitigating PPP through Covariance Matrix Simplification}
Since strong correlations within the data measurements are a necessary condition for PPP, one approach to mitigate this issue is to eliminate the non-diagonal terms in the data covariance matrix. While this explicitly avoids the PPP problem, it effectively invalidates the uncertainties reported by the experiments and creates challenges when results from multiple experiments are included in the model tuning exercise. However, in cases where only a single experimental result is included and there are alternative ways to estimate the uncertainties of the model parameters, this approach can yield decent results. For example, in Ref.~\cite{MicroBooNE:2021ccs} a series of iterative fits were performed to data with only diagonal uncertainties used, and post-fit model uncertainties were chosen to cover the range of best-fit model parameter values encountered throughout the iterative fitting process. It is worth noting that the referenced work also investigated fits using the covariance matrix transform described in the next section and found good agreement with the diagonal-uncertainties-only fits.

\subsection{Mitigating PPP through Covariance Matrix Transformation}
\begin{figure}[!htbp]
    \centering
    \includegraphics[width=1.0\linewidth]{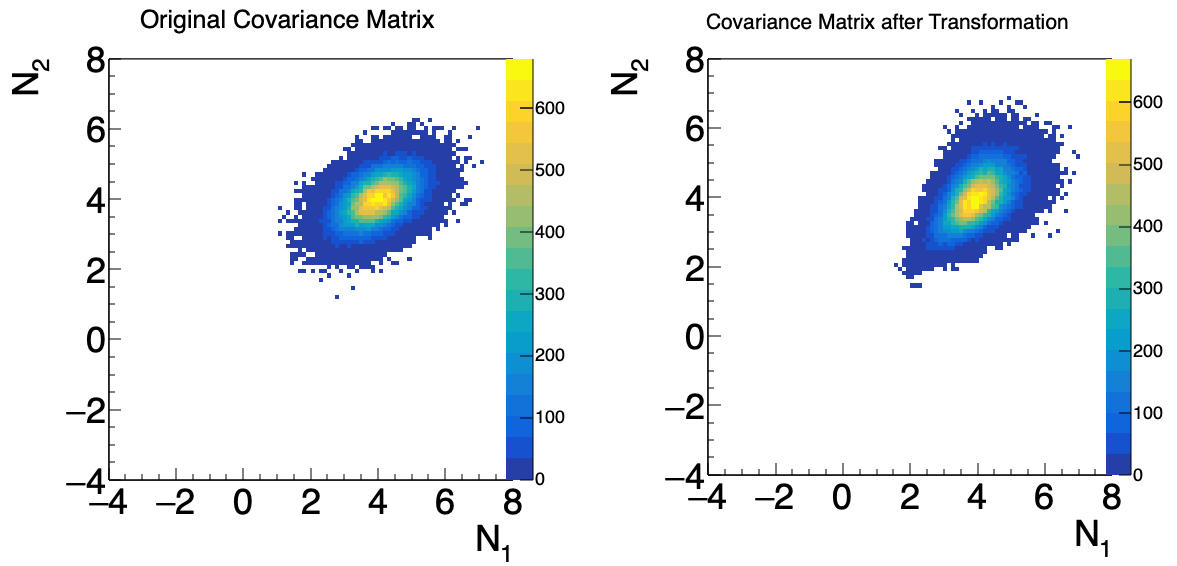}
    \caption{Comparison of data correlations for the original two-bin data measurement (left) and those after the non-linear norm-shape transformation (right).}
    \label{fig:cov_matrix_shape_norm}
\end{figure}

As illustrated in Ref.~\cite{Chakrani:2023htw}, another approach to addressing the PPP issue is to perform a non-linear norm-shape conversion on the data results and their covariance matrix. The new set of variables, $\mathcal{C} = \{C_1, \ldots, C_m\}$, are defined as follows:
\begin{equation}
C_i = f(n_i) = \begin{cases}
\alpha \frac{n_i}{\sum_k n_k}, & 1 \leq i \leq m - 1 \\
N_T = \sum_k n_k, & i = m
\end{cases}
\end{equation}
where $\alpha$ is a scale parameter that can be chosen, and $m$ is the dimension of the data measurement. Despite this transformation being non-linear, the covariance matrix of $\{C_1, \ldots, C_m\}$ can be estimated using the error propagation rule:
\begin{equation}
\text{Cov}[C] = J(f) \cdot \Sigma_{\text{data}} \cdot J(f)^T 
\end{equation}
where \(J(f)\) is the Jacobian of the non-linear transformation \(f\). The new covariance matrix 
is expressed as follows:
\begin{widetext}
\begin{equation}
   \text{Cov}[C] =  
    \begin{pmatrix}
     \frac{\alpha^2}{N_T^2} \left( \sigma_{i,j} - \frac{n_i}{N_T} \sum_l \sigma_{j,l} - \frac{n_j}{N_T} \sum_k \sigma_{k,i} + \frac{n_i n_j}{N_T^2} \sum_{kl} \sigma_{k,l} \right)& \rvline &
        \begin{matrix}
            \frac{\alpha}{N_T} \left(\sum_l \sigma_{1,l} - \frac{n_1}{N_T} \sum_{kl} \sigma_{k,l} \right)\\[1em]
            \vdots\\[1em]
            \frac{\alpha}{N_T} \left(\sum_l \sigma_{m-1,l} - \frac{n_{m-1}}{N_T} \sum_{kl} \sigma_{k,l} \right)
        \end{matrix}\\[1em]
        \hline
        \begin{matrix}
            \frac{\alpha}{N_T} \left(\sum_k \sigma_{k,1} - \frac{n_1}{N_T} \sum_{kl} \sigma_{k,l} \right) &
            \cdots &
            \frac{\alpha }{N_T} \left(\sum_k \sigma_{k,m-1} - \frac{n_{m-1}}{N_T} \sum_{kl} \sigma_{k,l} \right)
        \end{matrix}
        & \rvline & \sum_{kl} \sigma_{k,l}
    \end{pmatrix},
\end{equation}
\end{widetext}
where $\sigma_{kl}$ is an element of the data covariance matrix $\Sigma_{\text{data}}$. We reproduce the above equation to correct typos in Ref.~\cite{Chakrani:2023htw}. Figure~\ref{fig:cov_matrix_shape_norm} illustrates the transformation of the allowed measurement distributions in a simple two-bin measurement. While there is no information loss in this transformation, the non-linear nature of the transformation alters the correlations between the data points.
While this approach also avoids the PPP issue, it is not ideal because it applies a non-linear transformation that alters the correlations reported by the experimental collaboration. These correlations that should be preserved to maintain the integrity of the original analysis.

Furthermore, these two approaches discussed above primarily address the normalization issue identified in the original PPP phenomenon. As discussed in Sec.~\ref{sec:methodology}, data-model mismatches can also occur in higher-dimensional shape analyses involving non-normalization PPP-like issues. Ideally, a comprehensive approach would address both normalization and shape dimension mismatches inclusively. 

\subsection{Mitigating Inconsistencies between Data and Model through Quantile Mapping}
We propose a strategy to mitigate inconsistencies between data and model predictions, addressing not only PPP but also broader issues that can arise in such comparisons. Central to this approach is the investigation of bin-by-bin disagreements in the diagonalized space discussed earlier. This basis allows for the identification and resolution of inconsistencies that extend beyond overall normalization effects, as demonstrated in Sec.~\ref{sec:studies}. By adopting this comprehensive framework, we aim to provide a more robust solution for resolving data-model tensions.

A key challenge in fitting a model to data in the presence of improper treatments is managing the exaggerated discrepancies between measurement and prediction. Such discrepancies often manifest as excessively large local $\chi^{2}_{i}$ terms, which distort the overall $\chi^{2}$ minimization process and can lead to extreme or non-physical best-fit parameters. This issue arises because the covariance matrix $\Sigma_{D}$ is insufficient to describe the observed differences D under the influence of a distorting improper treatment.

To address this, we propose a procedure inspired by the process of Quantile Mapping (QM)~\cite{quantile_mapping}, which adjusts the uncertainties in the diagonalized data-minus-measurement covariance matrix $\Sigma_{D}$ . By enlarging the uncertainties for specific elements, the adjusted covariance matrix $\widetilde{\Sigma}_{D}$ becomes capable of accounting for the observed disagreements between measurement and prediction. This adjustment reduces the excessive contributions of individual bins to the overall $\chi^{2}$, mitigating the impact of improper treatments and enabling more reliable and meaningful fit results.

We provide a simple description of QM here, and leave a more rigorous derivation of the approach to the appendix. The central idea of the procedure is that when a model accurately describes a measurement, referred to here as the null hypothesis, the individual $\chi^{2}_{i}$ terms will be sampled from a $\chi^{2}$ distribution with one degree of freedom. These $n$ $\chi^{2}_{i}$ terms can then be arranged in increasing order and described by the cumulative distribution function (CDF) $g(x)$:
\begin{equation}
    g(x) = K(x)/n,
\end{equation}
where $K(x)$ counts the number of bins with a $\chi^{2}_{i}$ value less than or equal to $x$ over the domain $\mathbb{R} \geq 0$.

For a large number of bins, if the null hypothesis is true then the observed CDF will approximate the CDF of the $\chi^{2}$ distribution with one degree of freedom. If the observed $\chi^{2}_{i}$ terms are larger than expected, this would indicate that the corresponding uncertainties in $\Sigma_{D}$ are too small. Therefore, if the observed CDF does not match the expected distribution in this manner, the data uncertainty of individual bins can be enlarged to map each $\chi^{2}_{i}$ value onto the corresponding value for the CDF of a $\chi^{2}$ distribution with one degree of freedom. We recommend that uncertainties are only maintained or enlarged, but not shrunk in this manner, so that the QM treatment remains conservative in terms of its uncertainty estimation. This approach will yield an adjusted data-model comparison where the newly determined uncertainties are able to describe the observed differences.

Before concluding this section, it is important to emphasize that the proposed procedures are designed to mitigate, rather than fully eliminate, the negative impacts of improper treatments in data-model comparisons. While the underlying causes of these improper treatments remain unaddressed, these strategies focus on minimizing their adverse effects to enable meaningful model fitting. Notably, the quantile mapping approach provides a flexible framework for incorporating multiple measurements, even when some are affected by issues like PPP. By enlarging uncertainties for comparisons suspected of improper treatments, this method allows such measurements to be included in a joint fit in a conservative manner. This approach naturally under-emphasizes these problematic comparisons while prioritizing those with no suspected improper treatment, enhancing the robustness and reliability of the overall analysis. However, if QM is applied in cases where there is no improper treatment and only a moderate model inaccuracy, QM will decrease the capability of the measurement to inform legitimate model refinement. Therefore, it is important to only apply QM in cases where one is reasonably confident that an improper comparison or otherwise severe data-model mismatch is present and hampering successful fit attempts.

\section{Case Studies of Modeling Fitting}\label{sec:studies}

We present multiple case studies that illustrate types of potential mismatches and demonstrate the capabilities of the proposed methods in detecting and mitigating improper treatments. We compare published cross sections to model predictions from commonly used cross section event generators such as \texttt{GENIE} and \texttt{NEUT}, with a selection of post-fit parameter values shown in the appendix. In each case, the comparison to model predictions both before and after fitting, including calculated $\chi^{2}$ values, is made using the nominal model uncertainties that are described for each model prediction. In some cases, part or all of the model used to predict a cross section is the same as the model used to unfold the data measurement. This means there are potential correlations between the data and the predictions they are compared against that are not well captured in this analysis. However, these correlations are likely small, given that neutrino interaction uncertainties are sub-dominant in the cross section measurements in these case studies, in part because they are naturally suppressed in the unfolding process. Furthermore, many case studies highlight particular improper treatments by comparing model fits with and without artificially introducing the improper treatment of interest, and only find evidence of PPP when the desired improper treatment is artificially added. This demonstrates that the artificially added improper treatment is responsible for the PPP effect observed. Therefore, overall we do not believe that the potential correlations in neutrino interaction modeling between a data measurement and a model prediction play a significant role in these studies.

\subsection{Inconsistencies between Data and Models from Regularizations in Data Unfolding}\label{sec:reg_case}

The first case study examines the impact of regularization in unfolding on the quality of model fits, specifically when the effects of regularization are not properly accounted for in data-model comparisons. Unfolding methods are widely used to reconstruct event count distributions and produce cross section measurements in terms of true kinematics. Many of these methods, such as Wiener singular value decomposition (SVD)~\cite{Tang:2017rob} and D’Agostini unfolding~\cite{DAgostini:1994fjx}, rely on regularization to stabilize the unfolding process and suppress statistical fluctuations. However, regularization can introduce non-physical distortions into the unfolded data compared to an unregularized counterpart.

Methods like Wiener SVD address this issue by calculating a regularization matrix, $A_{C}$, which accounts for the effects of regularization. This process can be generalized to many other kinds of unfolding~\cite{gardiner2024} as well. Applying the regularization matrix to model predictions ensures that both the data and model predictions are treated consistently, avoiding the introduction of bias in their comparison. By contrast, failing to account for regularization effects creates mismatches between the unfolded data and model predictions, leading to unreliable fit results and potentially improper interpretations.

To illustrate the impact of regularization in data-model comparisons, we perform a series of model fittings to measurements reported by MicroBooNE in one, two, and three dimensions. These measurements all investigate inclusive muon neutrino charged-current ($\nu_{\mu}$CC) interactions on argon using very similar signal definitions, event selections, and unfolding methodologies. They include single- and double-differential cross sections $d\sigma/dE_{\mu}$ and $d^{2}\sigma/dE_{\mu}d\cos\theta_{\mu}$~\cite{PhysRevLett.133.041801} as well as the three-dimensional inclusive cross section $d^{2}\sigma(E_{\nu})/dP_{\mu}d\cos\theta_{\mu}$~\cite{MicroBooNE:2023foc}. For each measurement, model fits are attempted both with and without applying the regularization matrix, which represents realistic scenarios where the regularization matrix was not reported. In cases where the regularization matrix was omitted from the fit, it is also omitted from the $\chi^{2}$ calculation.

The \texttt{NUISANCE} package~\cite{Stowell:2016jfr} is used to compute model predictions for comparison with the reported cross sections. In this section we use a model based on \texttt{GENIE} v3.0.6 G18\_10a\_02\_11a, tuned to T2K data~\cite{MicroBooNE:2021ccs,T2K:2016jor}, and referred to as the \texttt{GENIE nominal} model. Uncertainties for each of the 55 model parameters follow the treatment outlined in the referenced tune~\cite{MicroBooNE:2021ccs}, with the exception of FrPiProd\_N and FrPiProd\_pi, which were omitted for technical limitations in the implementation. Using \texttt{NUISANCE} and the provided code framework~\cite{conditional_constraint_example}, for each measurement we construct both the central value prediction of the model and a covariance matrix that describes the overall uncertainty in the model prediction, including correlations between the measurement space and model parameters. In cases where we apply the regularization matrix, this is done before constructing the covariance matrix for the model uncertainty, meaning that the covariance is computed in the regularized space. Leveraging the covariance matrix formalism described in \autoref{sec:cov_matrix}, we perform a fit of the model to the data measurement, enabling a detailed investigation of the role of regularization in the unfolding process.

\begin{figure}[hbt!]
     \centering
     \includegraphics[clip,width=0.5\textwidth]{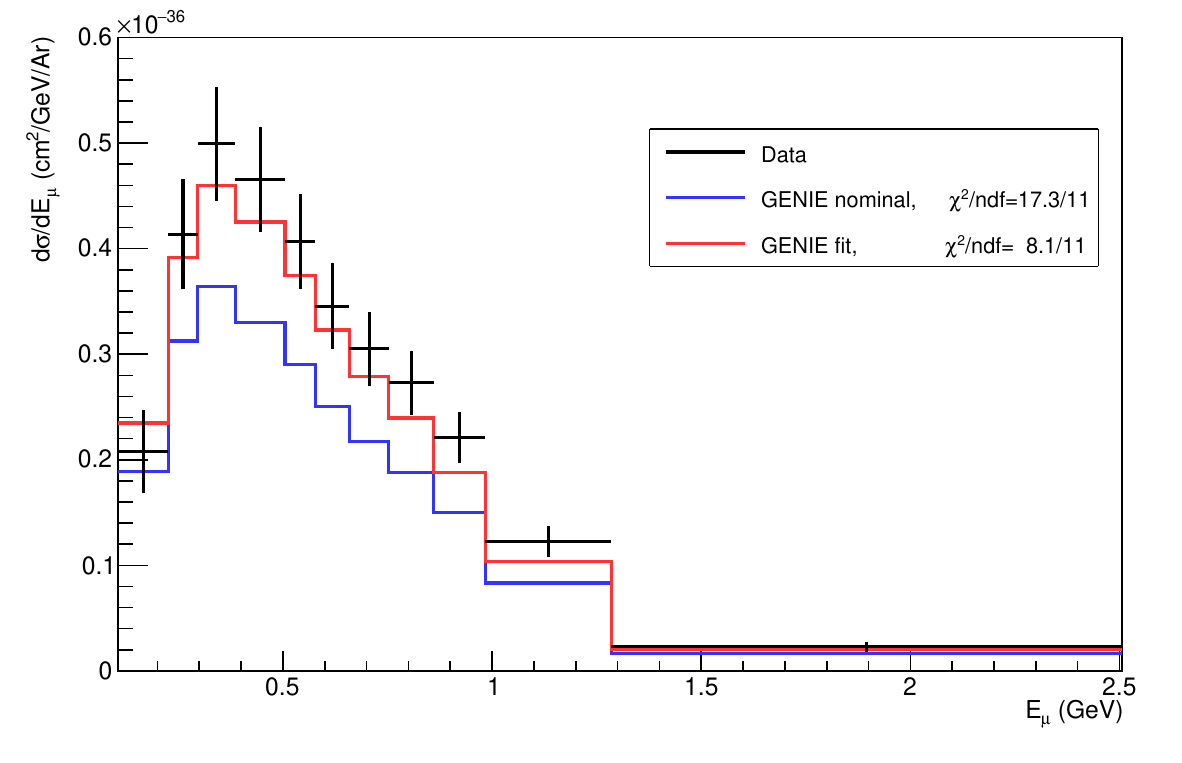}
     \put(-100,158){\normalsize{MicroBooNE Data}}
     \caption{Comparison of the measured MicroBooNE $d\sigma/dE_{\mu}$ cross section data (black) to the nominal \texttt{GENIE} model prediction with the regularization matrix applied (blue), and the fit result with the regularization matrix applied (red).}
    \label{fig:MicroBooNE_EMu}
\end{figure}

\begin{figure}[hbt!]
     \centering
     \includegraphics[clip,width=0.5\textwidth]{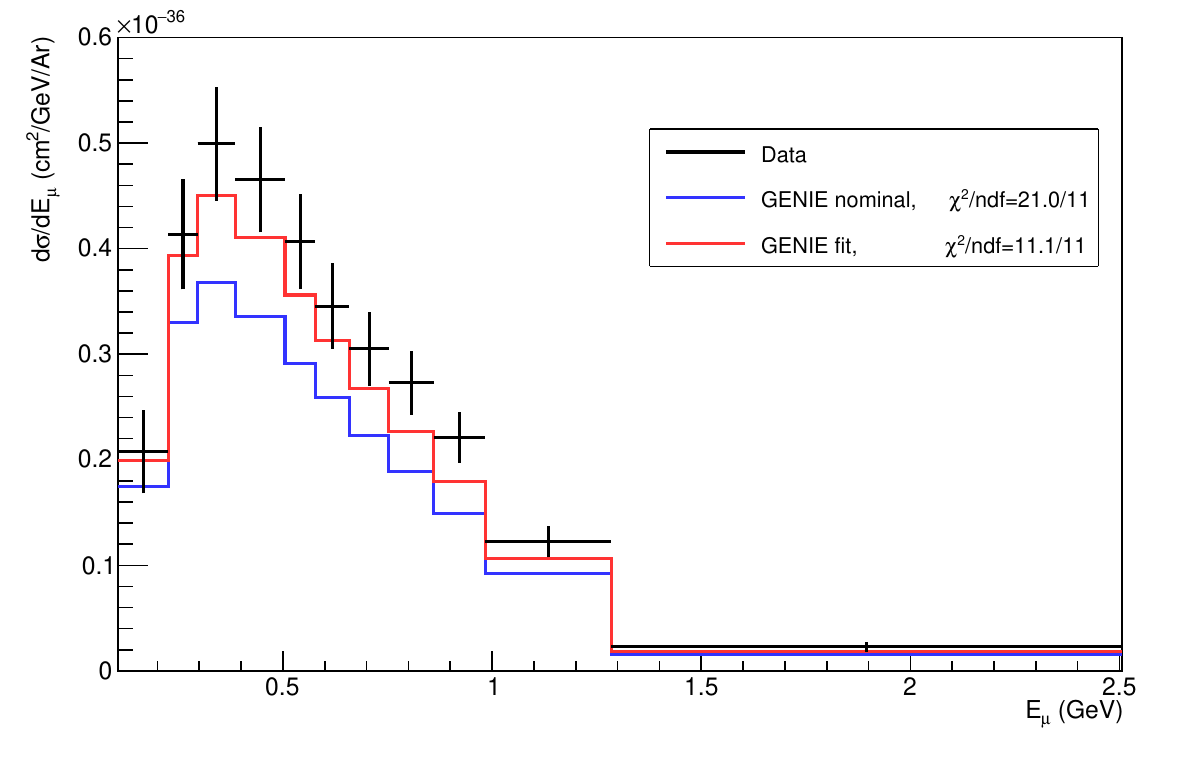}
     \put(-100,158){\normalsize{MicroBooNE Data}}
     \put(-85,95){\footnotesize{No $A_{C}$ Matrix}}
     \caption{Comparison of the measured MicroBooNE $d\sigma/dE_{\mu}$ cross section data (black) to the nominal \texttt{GENIE} model prediction without applying the regularization matrix (blue), and the fit result without applying the regularization matrix (red).}
    \label{fig:MicroBooNE_EMu_noac}
\end{figure}

First we investigate the quality of fits performed to the single-differential cross section measurement $d\sigma/dE_{\mu}$, shown in \autoref{fig:MicroBooNE_EMu} and \autoref{fig:MicroBooNE_EMu_noac}. When the regularization matrix is applied, the post-fit model prediction achieves a $\chi^{2}/\mathrm{ndf}$ of 8.1/11 with an associated \textit{p}-value of 0.70, indicating a good agreement between the model and data without any clear evidence of a normalization issue indicative of PPP. When the regularization matrix is omitted, the fit quality is degraded with a $\chi^{2}/\mathrm{ndf}$ of 11.1/11 and an associated \textit{p}-value of 0.43. Although the quality of the fit has worsened, there is no significant sign of a PPP issue. This indicates that omission of the regularization matrix will harm the quality of model fit attempts, but may not necessarily present a detectable PPP issue.

\begin{figure}[hbt!]
     \centering
     \includegraphics[clip,width=0.5\textwidth]{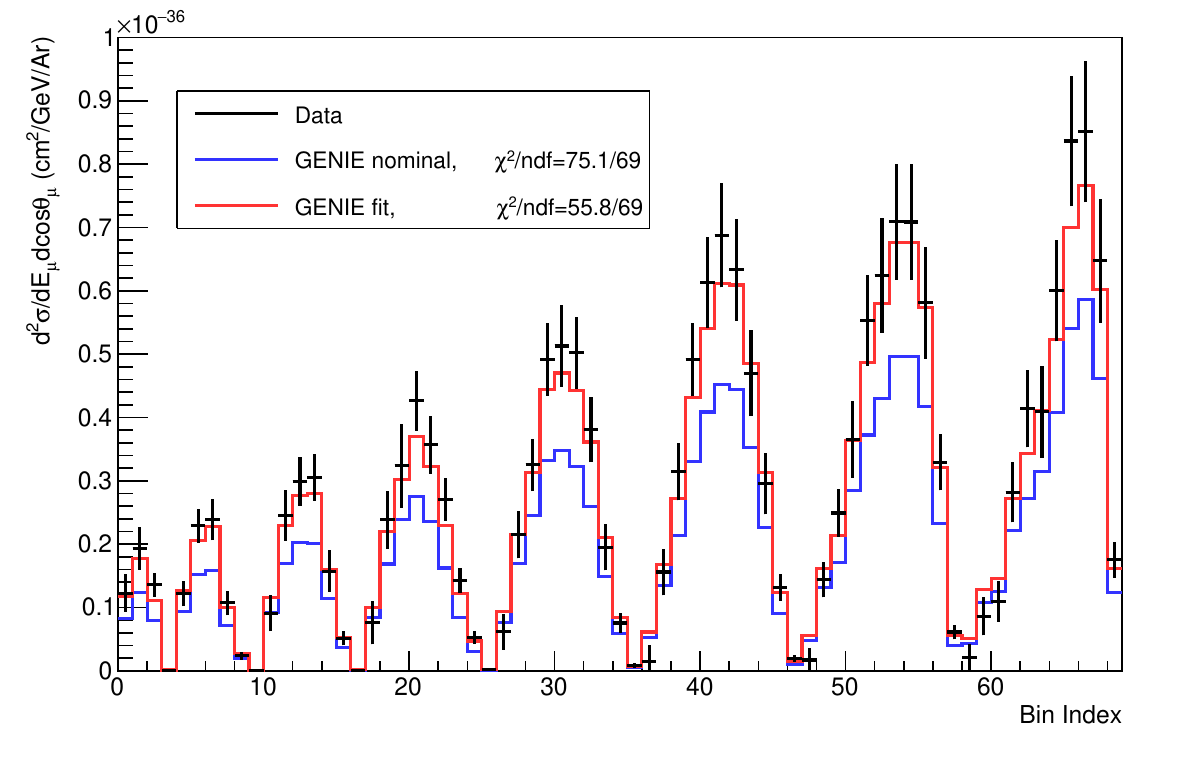}
     \put(-100,158){\normalsize{MicroBooNE Data}}
     \caption{Comparison of the measured MicroBooNE $d^{2}\sigma/dE_{\mu}d\cos\theta_{\mu}$ cross section data (black) to the nominal \texttt{GENIE} model prediction with the regularization matrix applied (blue), and the fit result with the regularization matrix applied (red). The x-axis represents the bin index.}
    \label{fig:MicroBooNE_EMuCosThetaMu}
\end{figure}

\begin{figure}[hbt!]
     \centering
     \includegraphics[clip,width=0.5\textwidth]{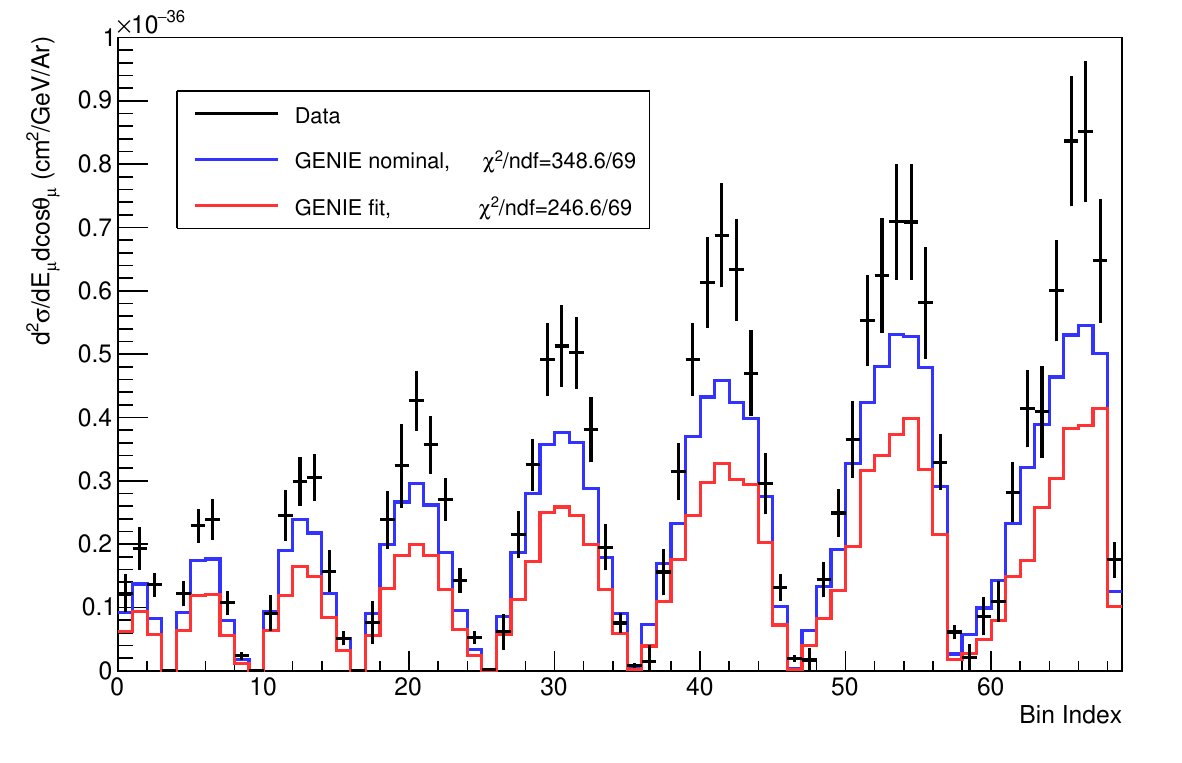}
     \put(-100,158){\normalsize{MicroBooNE Data}}
     \put(-200,105){\footnotesize{No $A_{C}$ Matrix}}
     \caption{Comparison of the measured MicroBooNE $d^{2}\sigma/dE_{\mu}d\cos\theta_{\mu}$ cross section data (black) to the nominal \texttt{GENIE} model prediction without applying the regularization matrix (blue), and the fit result without applying the regularization matrix (red). The x-axis represents the bin index. The absence of the regularization matrix leads to significant normalization discrepancies.}
    \label{fig:MicroBooNE_EMuCosThetaMu_noac}
\end{figure}

Next we investigate the quality of fits performed to the double-differential cross section measurement $d^{2}\sigma/dE_{\mu}d\cos\theta_{\mu}$, shown in \autoref{fig:MicroBooNE_EMuCosThetaMu} and \autoref{fig:MicroBooNE_EMuCosThetaMu_noac}. When the regularization matrix is applied, the post-fit model prediction achieves a $\chi^{2}/\mathrm{ndf}$ of 55.8/69 with an associated \textit{p}-value of 0.87. Furthermore, an analysis of individual $\chi^{2}_{i}$ terms in the diagonalized covariance matrix basis reveals a largest individual $\chi^{2}_{i}$ value of 6.2, corresponding to a \textit{p}-value of 0.59 after applying the \v{S}id\'{a}k correction~\cite{Sidak:1967} to control the family-wise error rate. These comparisons all indicate a good agreement between the model and data without any clear evidence of a normalization issue indicative of PPP.

When the regularization matrix is omitted, shown in \autoref{fig:MicroBooNE_EMuCosThetaMu_noac}, the fit quality is significantly degraded with a clear normalization disagreement. The poor agreement between model and data is also apparent in the global $\chi^{2}/\mathrm{ndf}$, which shrinks from 348.6/69 from under the nominal model to 246.6/69 for the post-fit model, corresponding to an extremely low \textit{p}-value of $9 \times 10^{-22}$. Additionally, the most extreme individual $\chi^{2}_{i}$ term reaches 48, with a \v{S}id\'{a}k-corrected \textit{p}-value of $4 \times 10^{-12}$. A detailed examination of individual $\chi^{2}_{i}$ terms suggests the presence of non-normalization PPP issues in the fit. Specifically, many DoF exhibit noticeably larger data-model tension through enlarged $\chi^{2}_{i}$ after fitting. This issue disappears when the regularization matrix is applied to the model prediction before fitting, indicating that it is directly related to the improper treatment and a symptom of non-normalization PPP. Since the fit tensions extend beyond the normalization DoF, PPP-mitigation strategies limited to separately addressing normalization and shape are insufficient to resolve complex non-normalization PPP issues.

\begin{figure*}[hbt!]
     \centering
     \includegraphics[clip,width=0.95\textwidth]{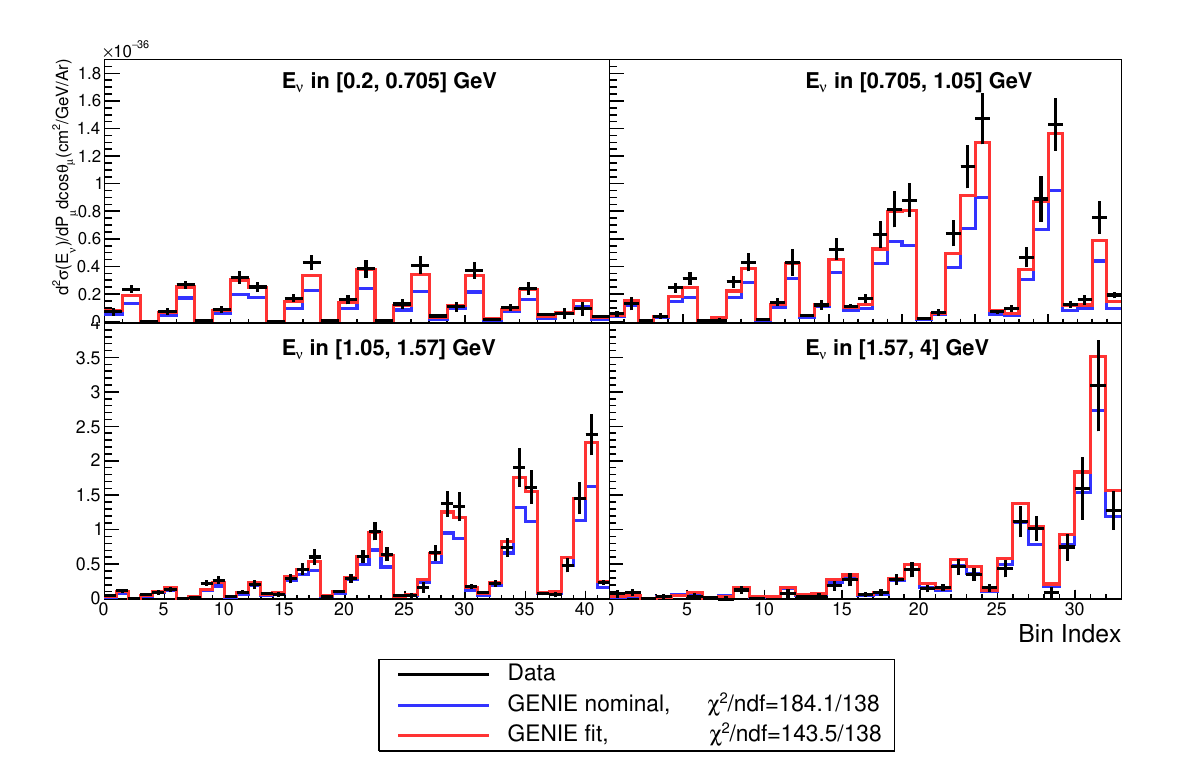}
     \put(-110,290){\normalsize{MicroBooNE Data}}
     \put(-451, 176){\color{white}{\rule{0.3cm}{0.3cm}}}
     \put(-237,  57){\color{white}{\rule{0.3cm}{0.3cm}}}
     \caption{Comparison of the measured MicroBooNE $d^{2}\sigma(E_{\nu})/dP_{\mu}d\cos\theta_{\mu}$ cross section data (black) to the nominal \texttt{GENIE} model prediction (blue), and the fit result (red).  Each sub-figure corresponds to a different slice of neutrino energy, and the x-axis represents the bin index for the 2D differential cross section within each $E_{\nu}$ slice.}
    \label{fig:numuCC_inc_3d_ac}
\end{figure*}

\begin{figure*}[hbt!]
     \centering
     \includegraphics[clip,width=0.95\textwidth]{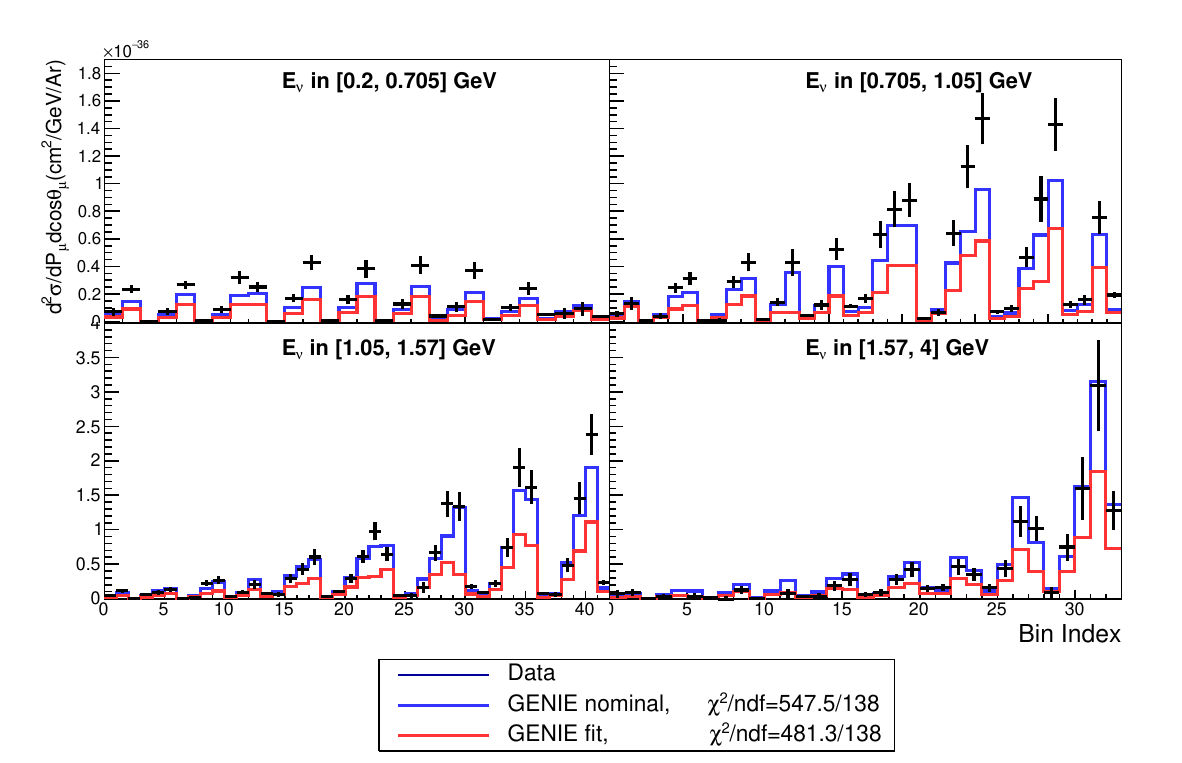}
     \put(-110,290){\normalsize{MicroBooNE Data}}
     \put(-451, 176){\color{white}{\rule{0.3cm}{0.3cm}}}
     \put(-237,  57){\color{white}{\rule{0.3cm}{0.3cm}}}
     \put(-321,  35){\color{black}{\rule{1.3cm}{0.05cm}}}
     \put(-265,-8){\normalsize{No $A_{C}$ Matrix}}
     \caption{Comparison of the measured MicroBooNE $d^{2}\sigma(E_{\nu})/dP_{\mu}d\cos\theta_{\mu}$ cross section data (black) to the nominal \texttt{GENIE} model prediction without applying the regularization matrix (blue), and the fit result without applying the regularization matrix (red).  Each sub-figure corresponds to a different slice of neutrino energy, and the x-axis represents the bin index for the 2D differential cross section within each $E_{\nu}$ slice. The absence of the regularization matrix leads to significant normalization discrepancies.}
    \label{fig:numuCC_inc_3d}
\end{figure*}

Finally, we investigate the quality of fits performed to the three-dimensional cross section measurement $d^{2}\sigma(E_{\nu})/dP_{\mu}d\cos\theta_{\mu}$, shown in \autoref{fig:numuCC_inc_3d_ac} and \autoref{fig:numuCC_inc_3d}. When the regularization matrix is applied, the post-fit model prediction achieves a $\chi^{2}/\mathrm{ndf}$ of 144.96/138 with an associated \textit{p}-value of 0.325. An analysis of individual $\chi^{2}_{i}$ terms in the diagonalized covariance matrix basis reveals a largest individual $\chi^{2}_{i}$ value of 8.95, corresponding to a \textit{p}-value of 0.319 after applying the \v{S}id\'{a}k correction. As with double-differential measurement, these comparisons all indicate a good agreement between the model and data without any clear evidence of a normalization issue indicative of PPP.

Again, when the regularization matrix is omitted, shown in \autoref{fig:numuCC_inc_3d}, the fit quality is significantly degraded with a clear normalization disagreement. The poor agreement between model and data is also apparent in the global $\chi^{2}/\mathrm{ndf}$ value of 616.77/138, corresponding to an extremely low \textit{p}-value of $4 \times 10^{-19}$. Additionally, the most extreme individual $\chi^{2}_{i}$ term reaches 52.76, with a \v{S}id\'{a}k-corrected \textit{p}-value of $10^{-11}$. Examination of individual $\chi^{2}_{i}$ terms shows non-normalization PPP issues through the existence of numerous DoF with significantly increased data-model tensions after fitting. As seen previously, the outliers disappear when the regularization matrix is applied to the model prediction before fit and comparison, demonstrating that they are driven by the improper treatment and a symptom of non-normalization PPP.

This case study highlights that omitting the regularization matrix, which arises from the data unfolding process, implicitly treats the unfolded results as truth-level quantities. This assumption introduces significant challenges in data-model comparisons, including PPP-related issues, particularly evident in normalization discrepancies, as well as broader inconsistencies between the data and model predictions. Without the regularization matrix, the regularization-induced distortions are not properly accounted for, which can lead to exaggerated penalties in the $\chi^{2}$ metric. Consequently, the fit may prioritize addressing artificial mismatches over resolving physically meaningful disagreements, resulting in unreliable and non-physical fit parameters, with the normalization issue being a clear manifestation of PPP. The effect of the data-model mismatch varies across the measurements studied, suggesting that some measurements, such as those in higher dimensions, are more susceptible to a severe PPP issue situation than others. Additionally, the absence of the regularization matrix compromises the integrity of the comparison, as it prevents equal treatment of the data and model predictions, thereby undermining the reliability of the extracted results. Including the regularization matrix is therefore critical for achieving meaningful and consistent comparisons in unfolding-based analyses. In \autoref{sec:QM_case}, we will further build on this case study to demonstrate the application of the QM mitigation strategy.

\subsection{Inconsistencies between Data and Models from Real vs. Nominal Neutrino Flux}\label{subsec:flux}
The second case study investigates the mismatch arising from assuming a certain flux to average the unfolded cross section over. They take the form of the ``real flux", where mainly the detector effects are unfolded and the cross section is averaged over the actual unknown neutrino flux, or the ``nominal flux", where a further translation is made to report a nominal flux-averaged cross section directly. While both are entirely self-consistent approaches, there are important trade-offs to consider in either one. As highlighted in~\cite{Koch:2020oyf}, the cross section measurements that are extracted at the real flux are less model dependent because the observed events in the detector are directly related to the convolution of the real flux and cross section. However, comparisons to this cross section using various model predictions have to then assume a flux, which is typically the nominal neutrino flux. The differences between the two fluxes will therefore lead to inconsistencies in the comparison, in particular by not fully accounting for the uncertainties in the flux shape in the reported measurement and their correlation with the flux uncertainties in the theoretical prediction.
The latter approach—--extracting the cross section at the nominal flux—--circumvents the issue, but introduces additional model dependence. This potential bias, however, can be assessed and constrained prior to unfolding through the data-driven validation methodology introduced in Ref.~\cite{MicroBooNE:2024kwe}.

In the current context, we are mainly interested in the impact of these assumptions to model fits. We construct a toy study where we  use the MicroBooNE three-dimensional $\nu_\mu$CC cross section results~\cite{MicroBooNE:2023foc} but introduce artificial mismatches related to the flux when fitting to the \texttt{GENIE nominal} model. Since these three-dimensional cross sections were originally extracted at the nominal flux, we know exactly the input flux used in the measurement and are therefore able to induce two mismatches when comparing to the model: 
\begin{itemize}
\item We first construct the model prediction by averaging over an alternative neutrino flux and not the nominal. This alternative flux is derived to be a plausible realization of the MicroBooNE flux covariance matrix. In particular, the first principal component of the flux covariance matrix is used to construct the deviation from the nominal flux at a level of $2\sigma$, while the other principal components are not varied. This represents the situation where there is a systematic issue in the nominal flux prediction but is nevertheless plausible relative to its uncertainties. 
\item Secondly, to mimic the situation in the ``real flux" approach more closely, we remove the contribution of the flux-shape uncertainties in the measurement 
by subtracting the total flux covariance matrix and replacing it with a $10\%$ normalization uncertainty denoting the overall error in the integrated flux.
\end{itemize}

We then compare the model prediction based on this alternative flux and re-factored uncertainties to the reported MicroBooNE measurement in \autoref{fig:MicroBooNE_Flux} below. As can be seen, there is a clear PPP-like normalization disagreement in the best-fit prediction to the data, even if the overall $\chi^{2}$ is much improved compared to the nominal case. 

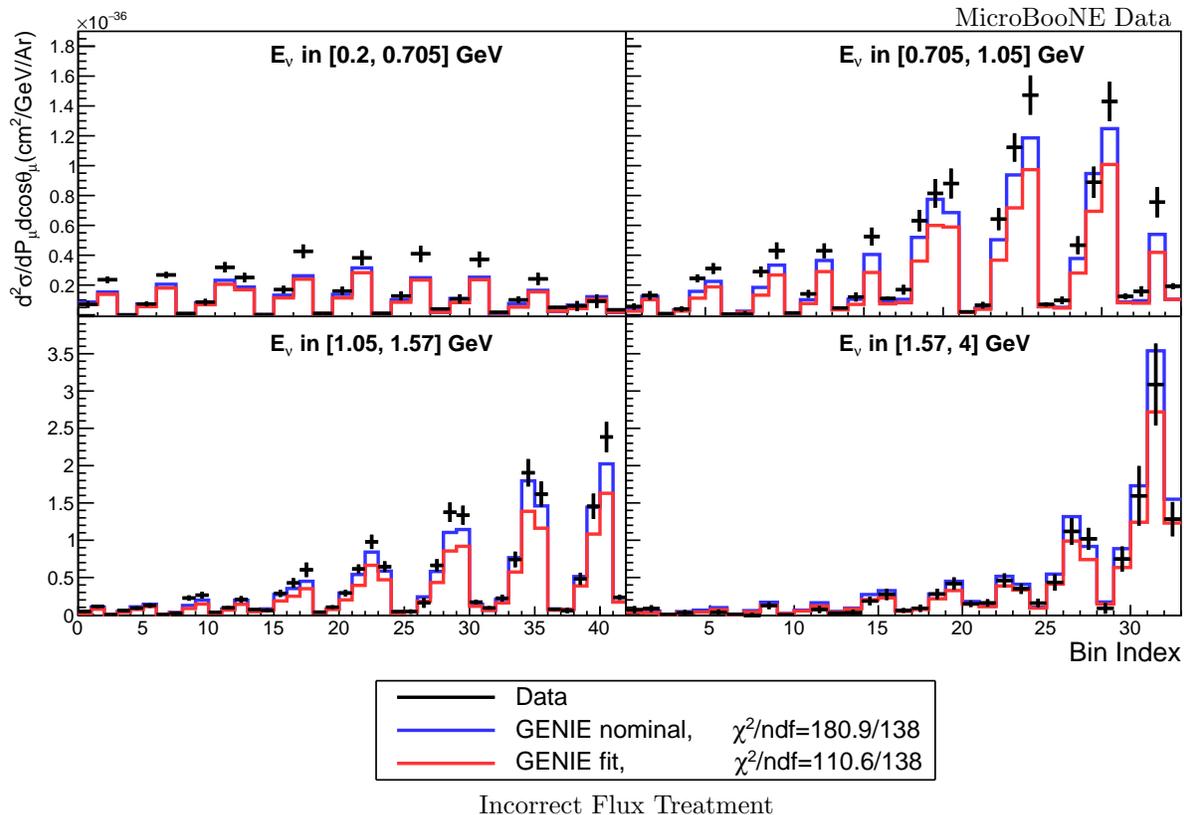
\begin{figure*}[hbt!]
     \centering
     \includegraphics[clip,width=0.95\textwidth]{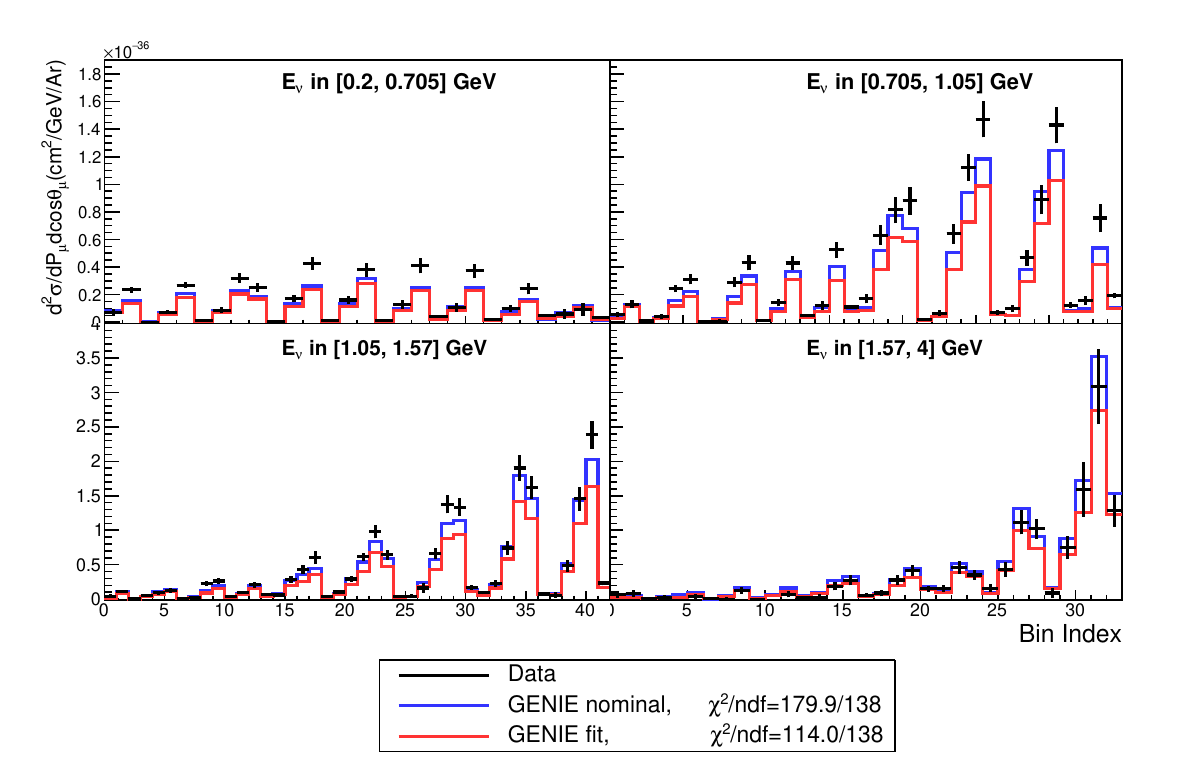}
     \put(-110,290){\normalsize{MicroBooNE Data}}
     \put(-451, 176){\color{white}{\rule{0.3cm}{0.3cm}}}
     \put(-237,  57){\color{white}{\rule{0.3cm}{0.3cm}}}
     \put(-290,-8){\normalsize{Incorrect Flux Treatment}}
     \caption{Comparison of the measured MicroBooNE $d^{2}\sigma(E_{\nu})/dP_{\mu}d\cos\theta_{\mu}$ cross section data (black) to the \texttt{GENIE} model prediction averaged over an alternative MicroBooNE flux (blue), and the corresponding fit result (red) to the data. The comparison as well as the fit is based on an alternative covariance matrix described above, which removes the effects of the flux-shape uncertainties.}
    \label{fig:MicroBooNE_Flux}
\end{figure*}

This case study highlights that mismatches between the real and nominal neutrino flux also introduce challenges in data-model comparisons, including PPP-related issues and broader inconsistencies. In practice, when cross sections are extracted at the real flux but model predictions are based on the nominal neutrino flux, systematic biases arise through the unmodeled correlations between measurement and prediction, leading to exaggerated penalties in the $\chi^{2}$ metric. This forces the fit to absorb flux-driven mismatches, potentially overshadowing meaningful physical discrepancies. As a result, the fit may yield unstable or nonphysical parameters, with normalization mismatches serving as a clear manifestation of the PPP effect. 

\subsection{Inconsistencies between Data and Models from Model Limitations}

The third case study investigates issues related to insufficient model coverage in data-model comparisons. To illustrate this point, we introduce the \texttt{NEUT 5.8.0} model~\cite{Hayato:2021heg} and the T2K 2020 $\nu_\mu$CC0$\pi$ measurement~\cite{T2K:2020jav} in addition to aforementioned \texttt{GENIE} model and MicroBooNE three-dimensional $\nu_\mu$CC cross section results. We use the unregularized version of the results reported by T2K, preventing the introduction of a PPP issue from any improper treatment related to regularization in the measurement.

The \texttt{NEUT} model incorporates the Spectral Function (SF) approach~\cite{Benhar:1994hw} based on Plane Wave Impulse Approximation (PWIA)~\cite{Ankowski:2014yfa}, which neglects final state interactions (FSI) via wave distortion, resulting in a significant overestimation of events at low $Q^2$.
In order to illustrate the impact of insufficient models, we restrict the model fitting analysis to six impactful parameters from the full set available for \texttt{NEUT}.  Among these parameters is the optical potential~\cite{Ankowski:2007uy}, which plays a critical role in describing nuclear effects. Other parameters are $M_A^{QE}$, one add-hoc high Q$^2$ normalizations for QE, a 2p2h normalization, and $M_A^{RES}$ and $C^A_{5}$ proposed by Graczyk and Sobczyk~\cite{SobczykGraczyk} for resonant production. By limiting the number of fitted parameters, the study focuses on how constrained model flexibility can exacerbate issues related to insufficient model coverage, particularly in regions of the phase space where the underlying physics is not fully captured.

\begin{figure*}[!htp]
    \centering
    \includegraphics[width=0.49\linewidth]{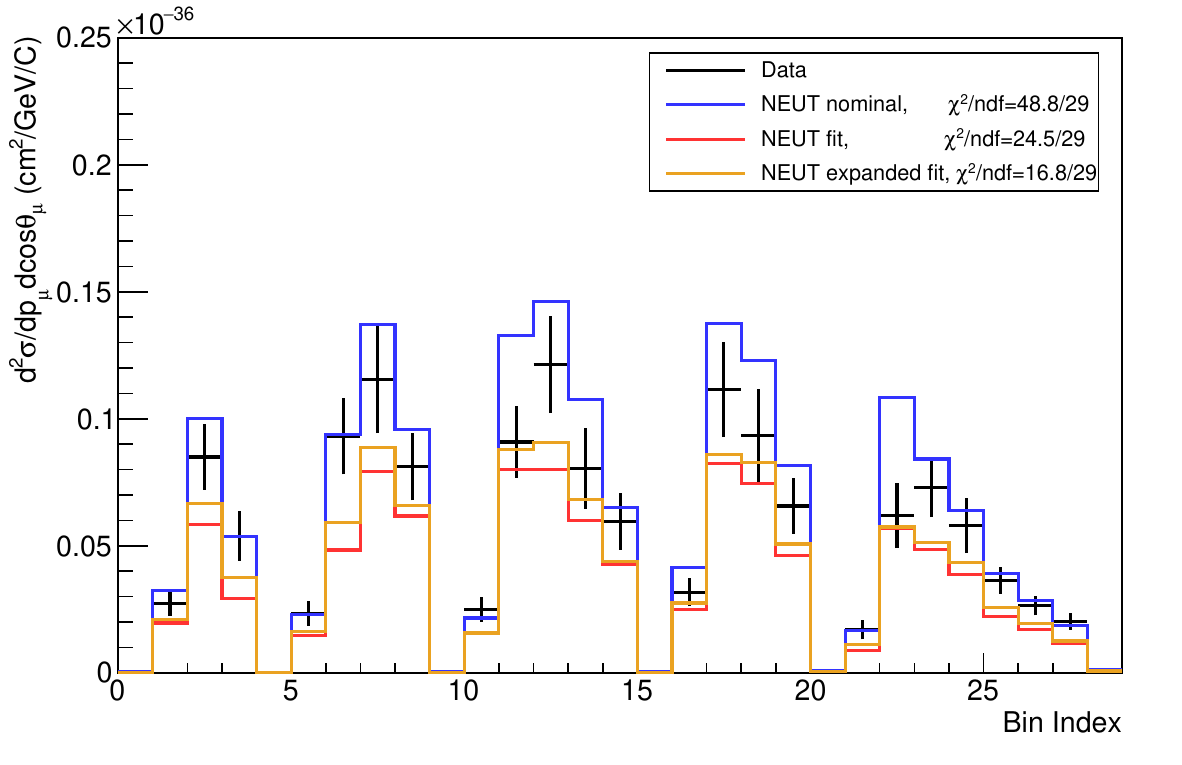}
    \put(-65, 155){\normalsize T2K Data}
    \put(-218.0,  19.0){\color{gray}{\rule{0.03cm}{3.35cm}}}
    \put(-188.6,  19.0){\color{gray}{\rule{0.03cm}{3.35cm}}}
    \put(-152.0,  19.0){\color{gray}{\rule{0.03cm}{3.35cm}}}
    \put(-108.0,  19.0){\color{gray}{\rule{0.03cm}{3.35cm}}}
    \put( -71.5,  19.0){\color{gray}{\rule{0.03cm}{3.35cm}}}
    \put(-207.0, 103.0){\footnotesize \boldmath{$\theta_{1}$}}
    \put(-173.0, 103.0){\footnotesize \boldmath{$\theta_{2}$}}
    \put(-130.9, 103.0){\footnotesize \boldmath{$\theta_{3}$}}
    \put( -92.0, 103.0){\footnotesize \boldmath{$\theta_{4}$}}
    \put( -50.0, 103.0){\footnotesize \boldmath{$\theta_{5}$}}
    \includegraphics[width=0.49\linewidth]{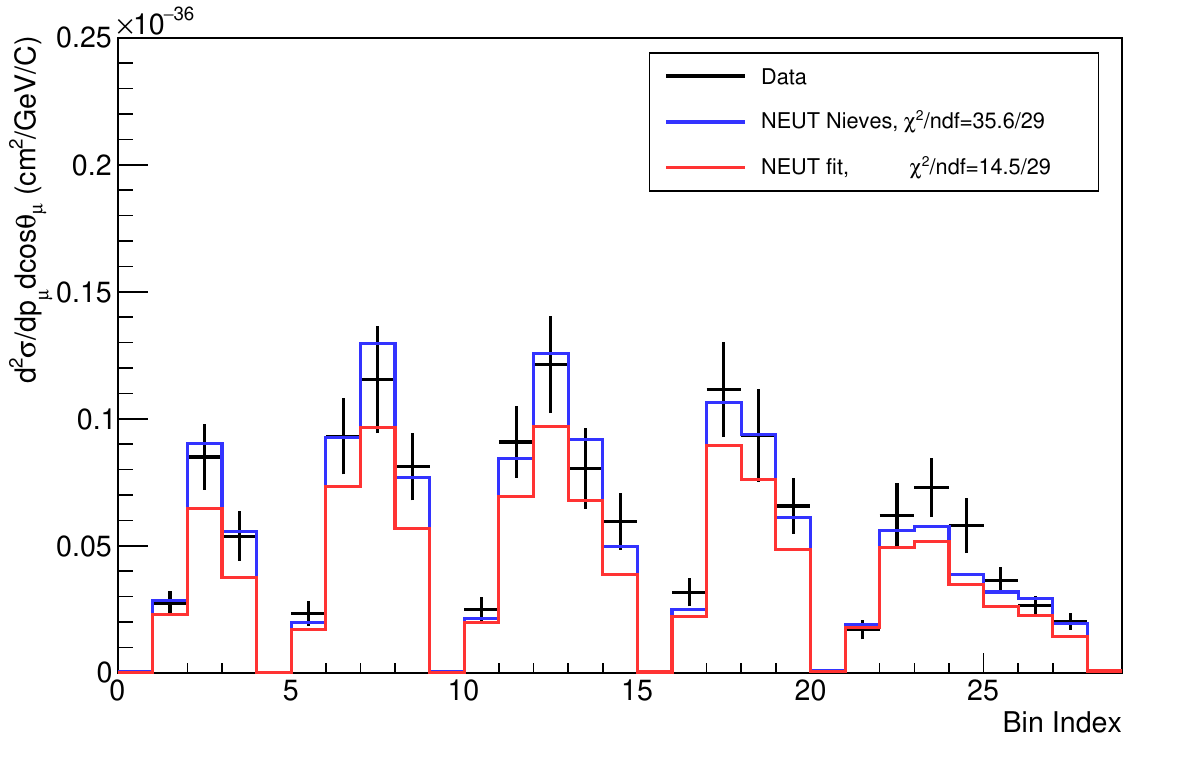}
    \put(-65, 155){\normalsize T2K Data}
    \put(-218.0,  19.0){\color{gray}{\rule{0.03cm}{3.35cm}}}
    \put(-188.6,  19.0){\color{gray}{\rule{0.03cm}{3.35cm}}}
    \put(-152.0,  19.0){\color{gray}{\rule{0.03cm}{3.35cm}}}
    \put(-108.0,  19.0){\color{gray}{\rule{0.03cm}{3.35cm}}}
    \put( -71.5,  19.0){\color{gray}{\rule{0.03cm}{3.35cm}}}
    \put(-207.0, 103.0){\footnotesize \boldmath{$\theta_{1}$}}
    \put(-173.0, 103.0){\footnotesize \boldmath{$\theta_{2}$}}
    \put(-130.9, 103.0){\footnotesize \boldmath{$\theta_{3}$}}
    \put( -92.0, 103.0){\footnotesize \boldmath{$\theta_{4}$}}
    \put( -50.0, 103.0){\footnotesize \boldmath{$\theta_{5}$}}
    \put(-380,-10){a)} 
    \put(-120,-10){b)} 
    \caption{ Comparison of the measured T2K CC0$\pi$ double differential cross section to a) the prior prediction of \texttt{NEUT} as well as two fits featuring six and nine free parameters labeled ``\texttt{NEUT} fit" and ``\texttt{NEUT} expanded fit", respectively, and b) the prior prediction of \texttt{NEUT} with Nieves model as well as a fit with six parameters. Each angle slice is separated with a vertical gray line and labeled, with a variable number of muon momentum bins spanning $[0,30]\,$GeV/c spanning each slice. }
    \label{fig:neutT2Kfit}
\end{figure*}
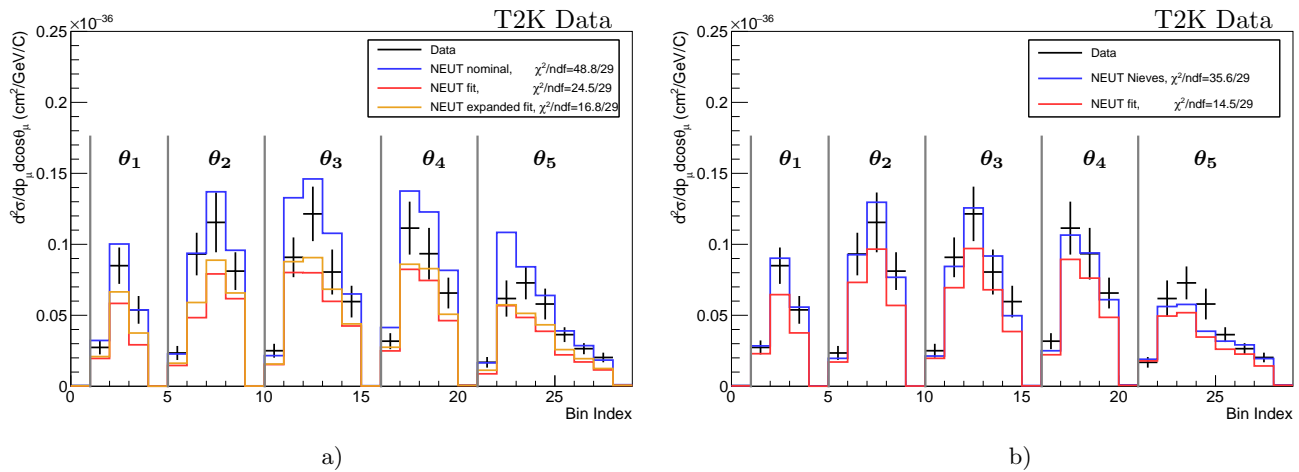

As shown in \autoref{fig:neutT2Kfit}a, the \texttt{NEUT} model introduced above overpredicts the data while the best-fit model, despite achieving a significantly lower $\chi^2$, under predicts the data, which is a behavior indicative of a PPP issue. To further investigate, the number of fit parameters are increased by introducing two additional ad-hoc normalization adjustments at high $Q^2$ and an additional freedom to account for Pauli blocking at low $Q^2$. This expanded model fit allows for slightly better post-fit data-model agreement, both in the $\chi^{2}$ and the normalization, however a significant normalization discrepancy indicative of a PPP issue still remains. This suggests that both model fits lack the freedom to explain the observed data-model disagreement.

To evaluate whether the observed misfit is due to insufficient model coverage or issues with the underlying base model, the \texttt{NEUT} model is modified through the inclusion of the Nieves 1p1h model~\cite{Nieves:2011pp} as a replacement for the Spectral Function (SF) model. As shown in Fig.~\ref{fig:neutT2Kfit}b, the \texttt{NEUT} model using Nieves 1p1h achieves a better data-model agreement than the previous \texttt{NEUT} model using a SF model, demonstrating an improved baseline agreement. This model is then fit using the same limited set of six parameters, not including the expanded fit parameters. The resulting post-fit $\chi^2$ decreases significantly, although there is still a notable normalization disagreement that arises from the fit attempt, indicating a PPP issue. However, the disagreement is less severe than in the \texttt{NEUT} model fits using the SF model, suggesting that the improved baseline comparison under the Nieves model helps reduce the PPP issue. Still, the limited number of fit parameters appears to be a contributing factor to the normalization discrepancy in all fit attempts.

It is worth noting the difficulty in real-world scenarios in attributing any observed PPP symptom to a specific cause. So far we have attempted to demonstrate the potential for model limitations to induce a PPP issue. However, it is possible that the observed PPP symptoms may be primarily driven by another issue than model limitations. For example, the T2K measurement used in these fit exercises is reported as a ``real-flux" measurement, while model predictions are averaged over a nominal flux prediction. As examined in detail in Sec.~\ref{subsec:flux}, this data-model mismatch can cause a PPP issue, meaning that it is possible that the observed PPP symptoms seen here may be driven more by this source or even another unknown source than by the model limitations that have been posited. Although perfect information is not available, it is possible to gain insight on the most likely reality from combining information across multiple studies. Relevant to this particular dataset, Ref.~\cite{Koch:2020oyf} demonstrated that the extra flux uncertainties generated from transforming an older T2K measurement~\cite{T2K:2016jor} with a similar phase space to a reference flux prediction~\cite{PhysRevD.87.012001} are small compared to the other uncertainties in the measurement. This suggests that in this case the impact of the improper treatment of comparing a ``real-flux" measurement to a ``nominal-flux" prediction may be small, reducing the likelihood that it is the primary cause of the observed PPP issues. Another way to test this hypothesis is to investigate additional model comparisons and fit attempts, and see whether they demonstrate similar PPP issues or not. If different outcomes are observed for different models, the flux mismatch hypothesis becomes disfavored.

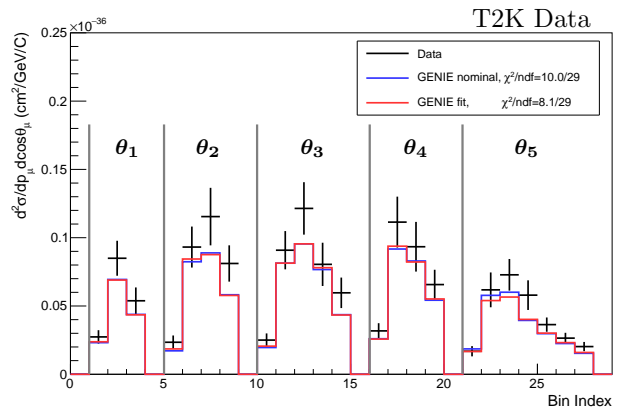
\begin{figure}[!htbp]
    \centering
    \includegraphics[width=0.98\linewidth]{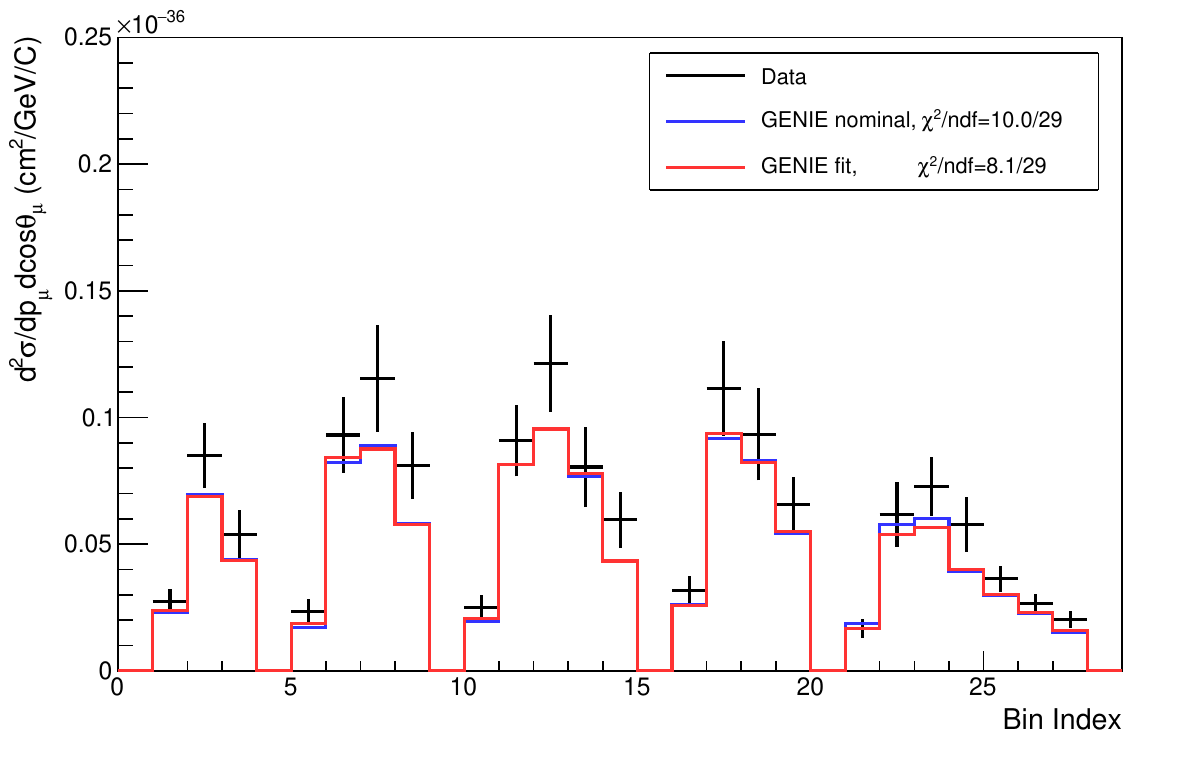}
    \put(-65, 150){\normalsize T2K Data}
    \put(-210.3,  18.8){\color{gray}{\rule{0.03cm}{3.32cm}}}
    \put(-182.0,  18.8){\color{gray}{\rule{0.03cm}{3.32cm}}}
    \put(-146.8,  18.8){\color{gray}{\rule{0.03cm}{3.32cm}}}
    \put(-104.2,  18.8){\color{gray}{\rule{0.03cm}{3.32cm}}}
    \put( -69.1,  18.8){\color{gray}{\rule{0.03cm}{3.32cm}}}
    \put(-200.0, 102.0){\footnotesize \boldmath{$\theta_{1}$}}
    \put(-169.5, 102.0){\footnotesize \boldmath{$\theta_{2}$}}
    \put(-130.0, 102.0){\footnotesize \boldmath{$\theta_{3}$}}
    \put( -91.0, 102.0){\footnotesize \boldmath{$\theta_{4}$}}
    \put( -49.0, 102.0){\footnotesize \boldmath{$\theta_{5}$}}

    \caption{ Comparison of the measured T2K CC0$\pi$ double differential cross section to \texttt{GENIE nominal} and tuned predictions before and after fitting with 4 free parameters. }
    \label{fig:genieT2K}
\end{figure}

Following this line of reasoning, we examine the comparison between the T2K data and the \texttt{GENIE nominal} and tuned predictions to the dataset, shown in \autoref{fig:genieT2K}. The latter prediction was tuned using four of the model's parameters: MaCCQE, NormCCMEC, RPA\_CCQE, and XSecShape\_CCMEC. The MaCCQE parameter adjusts the value of the axial mass in the dipole parameterization of the CCQE axial-vector form factor. The NormCCMEC parameter adjusts the normalization of the cross section of meson exchange current (MEC) interactions, which are the dominant type of charged-current 2-particle, 2-hole (CC2p2h) interactions in G18. The RPA\_CCQE parameter adjusts the strength of the suppression of the CCQE cross section at low Q$^2$ due to nucleon-nucleon long range correlations. The XSecShape\_CCMEC parameter adjusts the shape of the CC2p2h cross section between the Valencia model and the Empirical model, as described in Ref.~\cite{MicroBooNE:2021ccs}.

Comparing the nominal \texttt{GENIE} and \texttt{NEUT} predictions in Fig.~\ref{fig:genieT2K} and Fig.~\ref{fig:neutT2Kfit}, respectively, the nominal \texttt{GENIE} model prediction achieves much better agreement with the data, with a much lower $\chi^{2}/\mathrm{ndf}$. Additionally, the nominal \texttt{GENIE} model under-predicts the data, in contrast with the nominal \texttt{NEUT} model prediction which over-predicts the data. After fitting the \texttt{GENIE} model to the data, the $\chi^{2}/\mathrm{ndf}$ only improves slightly, with a minimal visual shift in the central value prediction. This may reflect a limitation in the fitted parameters in addressing the data-model discrepancy or may result from the fact that the nominal \texttt{GENIE} model prediction already achieved good agreement with the data, with a $\chi^{2}/\mathrm{ndf}$ well below unity. Either way, the overall normalization of the model prediction does not worsen after fitting, meaning the visible sign of a PPP issue that was present in each \texttt{NEUT} model fit is not reproduced when fitting the \texttt{GENIE} model to the same data set. This further supports the view that model limitations, and not flux issues, are the primary cause of the PPP issue observed in fit attempts with the \texttt{NEUT} model. Integrated cross sections for each model comparison to the T2K measurement are shown in Table~\ref{table:total_xs}. Comparing integrated cross section predictions between model versions demonstrates the presence of PPP normalization issues that were visually identified in Fig.~\ref{fig:neutT2Kfit}.

\begin{table}[]
    \centering
    \caption{Integrated cross sections per carbon atom for the T2K measurement~\cite{T2K:2020jav} and model predictions from \texttt{NEUT} and \texttt{GENIE}.}
    \begin{tabular}{|l|c|}
    \hline
    Model               & $\sigma (10^{-38}\mathrm{cm}^{2}/C)$ \\
    \hline
    T2K Measurement     & 5.64 $\pm$ 0.72 \\
    \texttt{NEUT} nominal        & 6.27 \\
    \texttt{NEUT} fit            & 3.54 \\
    \texttt{NEUT} expanded fit   & 4.13 \\
    \texttt{NEUT} Nieves         & 5.73 \\
    \texttt{NEUT} Nieves fit     & 4.21 \\
    \hline
    \texttt{GENIE} nominal       & 3.00 \\
    \texttt{GENIE} fit           & 3.02 \\
    \hline
    \end{tabular}
    \label{table:total_xs}
\end{table}

\begin{figure}[!htbp]
    \centering
    \includegraphics[width=0.98\linewidth]{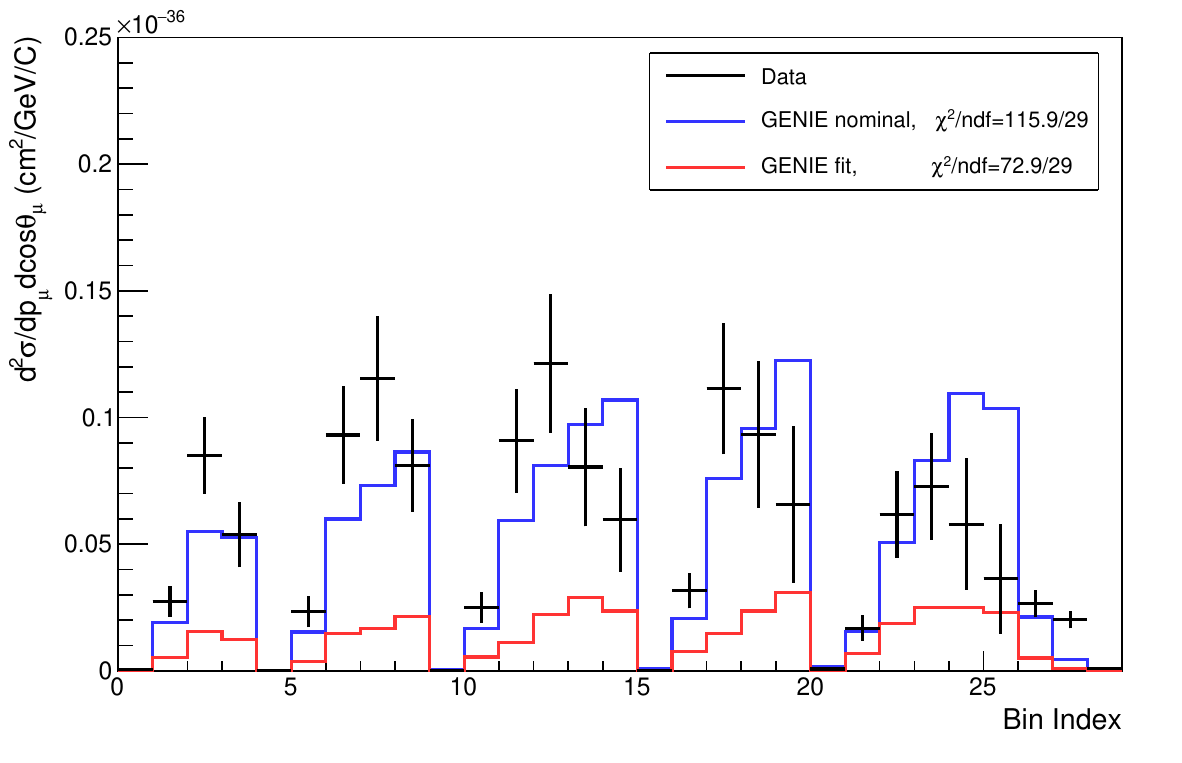}
    \put(-65, 150){\normalsize T2K Data}
    \put(-103,112){\tiny{Incorrectly Modeled Target}}
    \put(-210.3,  18.8){\color{gray}{\rule{0.03cm}{3.32cm}}}
    \put(-182.0,  18.8){\color{gray}{\rule{0.03cm}{3.32cm}}}
    \put(-146.8,  18.8){\color{gray}{\rule{0.03cm}{3.32cm}}}
    \put(-104.2,  18.8){\color{gray}{\rule{0.03cm}{3.2cm}}}
    \put( -69.1,  18.8){\color{gray}{\rule{0.03cm}{3.2cm}}}
    \put(-200.0, 102.0){\footnotesize \boldmath{$\theta_{1}$}}
    \put(-169.5, 102.0){\footnotesize \boldmath{$\theta_{2}$}}
    \put(-130.0, 102.0){\footnotesize \boldmath{$\theta_{3}$}}
    \put( -91.0, 102.0){\footnotesize \boldmath{$\theta_{4}$}}
    \put( -49.0, 102.0){\footnotesize \boldmath{$\theta_{5}$}}
    \caption{Comparison of the measured T2K CC0$\pi$ double differential cross section to the \texttt{GENIE G18\_10a\_02\_11a} model prediction erroneously calculated assuming an argon target. Both data and model cross sections are plotted per twelve nucleons (carbon). }
    \label{fig:genie_T2K_Ar}
\end{figure}
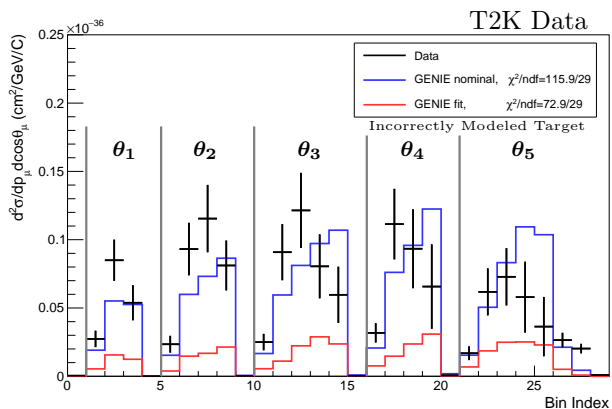

To directly illustrate how mismatches between data and model can trigger a PPP effect, we further consider the artificial model deficiency that arises when comparing a model prediction on argon to experimental cross section data from T2K, which is measured on a carbon target. The comparison on T2K carbon data is made in \autoref{fig:genie_T2K_Ar} using the \texttt{GENIE} model but erroneously computed under the assumption of an argon target. In contrast with the good model agreement found in \autoref{fig:genieT2K}, now there is significant disagreement seen between nominal model prediction and the T2K measurement. Given this improper treatment, the fit is no longer successful and shows clear signs of PPP through a significant normalization deficit.

\begin{figure*}[!htbp]
    \centering
    \includegraphics[width=0.95\textwidth]{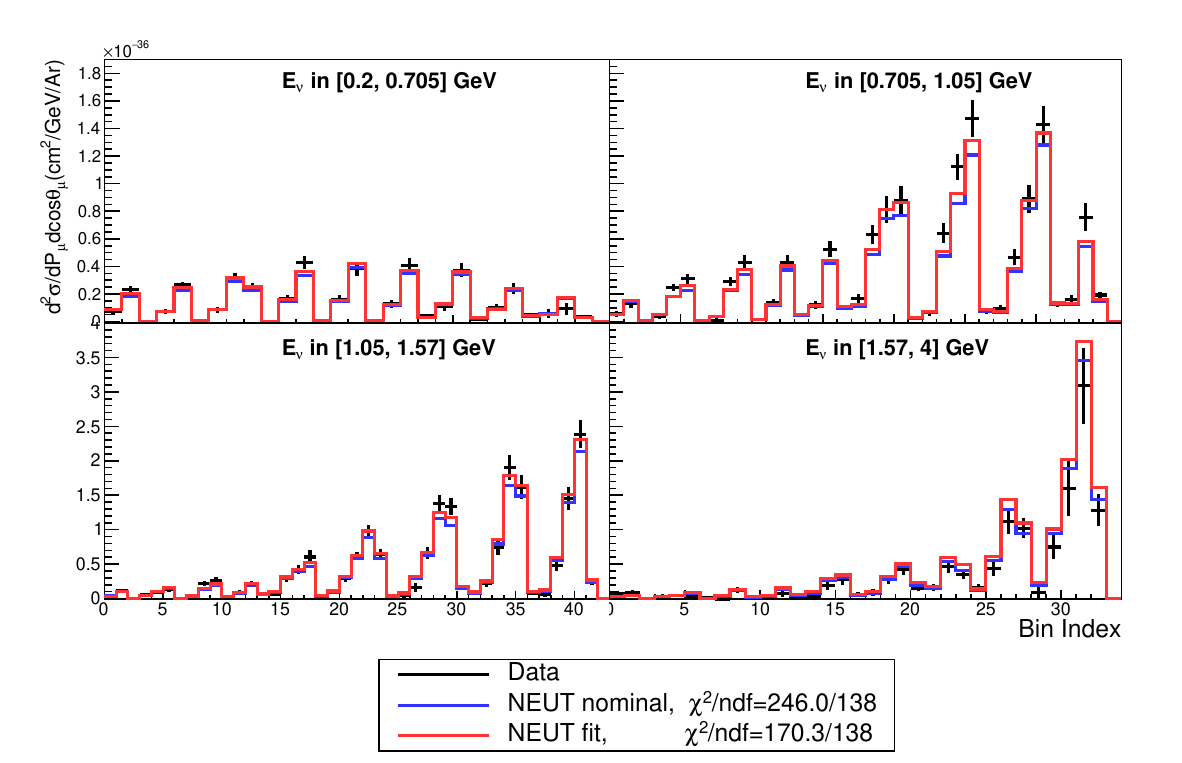}
     \put(-110,290){\normalsize{MicroBooNE Data}}
     \put(-451, 176){\color{white}{\rule{0.3cm}{0.3cm}}}
     \put(-237,  57){\color{white}{\rule{0.3cm}{0.3cm}}}
    \caption{Comparison of the measured MicroBooNE $d^{2}\sigma(E_{\nu})/dP_{\mu}d\cos\theta_{\mu}$ cross section to the \texttt{NEUT} model predictions before (blue) and after (red) fitting with six free parameters.}
    \label{fig:neutMicroBooNE}
\end{figure*}

While the six-parameter NEUT model exhibits the PPP effect when fitting the T2K data, it is also valuable to explore its performance on the three-dimensional $\nu_\mu$CC MicroBooNE data introduced in the previous section. 
Since NEUT does not include an SF for argon, the Nieves model has been used.  \autoref{fig:neutMicroBooNE} presents the results of this fit. Interestingly, while the $\chi^2$ value is not particularly low, there is no evident normalization discrepancy and no clear manifestation of a PPP effect. This outcome underscores an important nuance: although insufficient models can give rise to PPP, its emergence depends on the specific interplay between the data and the model. PPP is therefore not an inevitable consequence of model limitations, but rather arises under particular conditions where mismatches align in a way that accentuates the effect.

\subsection{Illustrating the QM Mitigation Strategy}\label{sec:QM_case}

In this section, we illustrate the quantile mapping (QM) mitigation strategy using the case study where a fit is attempted without applying the regularization matrix $A_C$ to model predictions for the double-differential cross section $d^{2}\sigma/dE_{\mu}d\cos\theta_{\mu}$. This scenario, which can occur if the regularization matrix is not reported, results in significantly degraded fit performance, as demonstrated in \autoref{sec:reg_case}. These examples highlight the challenges posed by regularization-induced distortions and serve as a basis for demonstrating how the QM approach can effectively address these issues.

\begin{figure}[hbt!]
     \centering
     \includegraphics[trim={0.0cm 0.0cm 0.0cm 0.0cm},clip,width=0.45\textwidth]{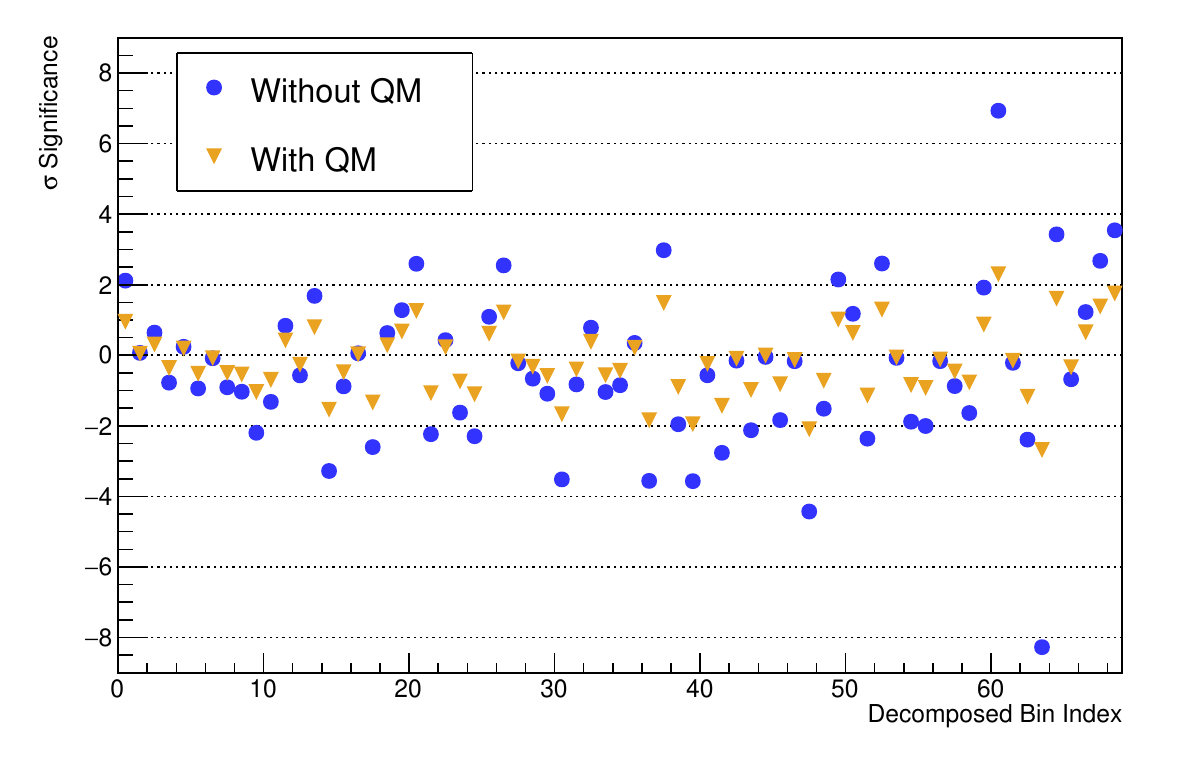}
     \put(-100, 142){\small MicroBooNE Data}
     \put(-195,101){\scriptsize{No $A_{C}$ Matrix}}
     \caption{Data-model $\sigma$ tension in each bin in the diagonalized covariance matrix basis for the measured MicroBooNE double-differential cross section $d^{2}\sigma/dE_{\mu}d\cos\theta_{\mu}$ before (blue) and after (green) applying QM.}
    \label{fig:nsigma}
\end{figure}

\begin{figure}[hbt!]
     \centering
     \includegraphics[clip,width=0.45\textwidth]{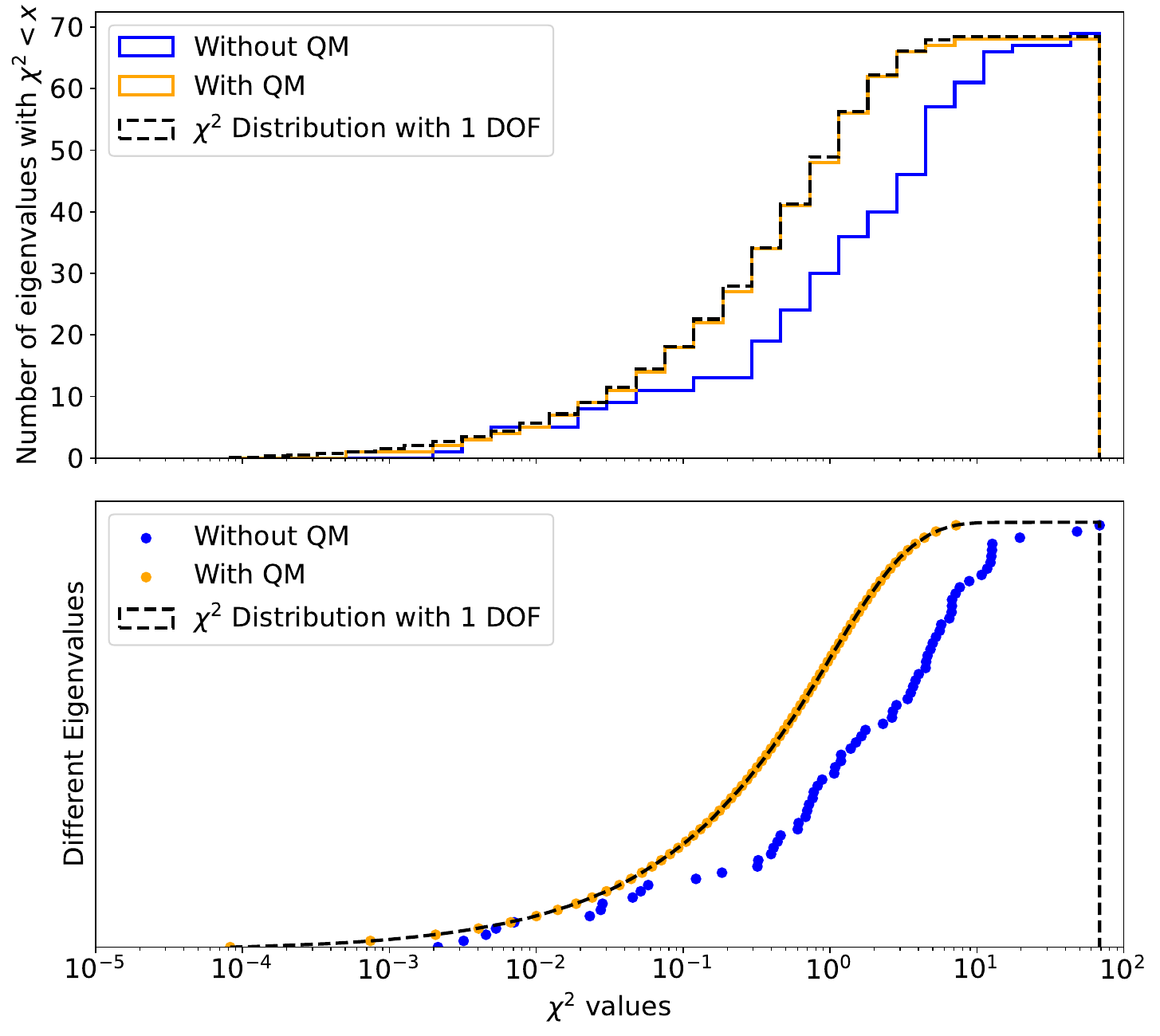}
     \put(-92, 204){\normalsize MicroBooNE Data}
     \put(-200,166){\scriptsize{No $A_{C}$ Matrix}}
     \caption{Distribution of individual $\chi^2_i$ values for the measured MicroBooNE double-differential cross section $d^{2}\sigma/dE_{\mu}d\cos\theta_{\mu}$ before and after applying the quantile mapping (QM) procedure. The distribution after QM aligns more closely with the expected $\chi^2$ distribution.}
    \label{fig:chi2_dist}
\end{figure}

As described in Sec.~\ref{sec:mitigation}, quantile mapping mitigates these PPP issues in the measurement by addressing the common symptom seen across many DoF. The mismatch between data and model prediction, magnified by a small covariance in some DoF, causes a naive fit attempt to focus on a few potentially insignificant diagonalized bins to the detriment of the model performance on other bins in the measurement space. The insufficiency of the nominal covariance matrix to describe the differences between data and nominal model prediction is shown in Fig.~\ref{fig:nsigma}. The blue circles represent the nominal data-model $\sigma$ tension in each bin in the diagonalized covariance matrix basis. The large number of points outside 2$\sigma$ provides a clear visual indication of the insufficiency of the nominal covariance matrix. This issue is addressed in the QM approach by enlarging the uncertainties in the diagonalized covariance matrix basis, restoring a suitable overall $\chi^{2}$ distribution as shown in Fig.~\ref{fig:chi2_dist}. As a result, a new at tempt at fitting the model to the data after using QM is able to give more appropriate weight to the disagreement within each DoF.

The improvement in the fitted model under QM can also be seen by comparing the parameter values between different fit procedures. Taking the parameter values achieved through fitting the model prediction with the $A_{C}$ matrix applied as the target outcome, we can compare the performances of the fits that omit the $A_{C}$ matrix, both with and without using QM. Focusing on the $\texttt{GENIE}$ parameters fitted in the $\texttt{MicroBooNE}$ tune~\cite{MicroBooNE:2021ccs}, the naive fit without use of the $A_{C}$ matrix or QM yields fit parameters in tension with their counterparts from the proper fit using the $A_{C}$ matrix at levels up to 6$\sigma$. In comparison, when QM is used tensions all drop to 2$\sigma$ or lower, demonstrating how QM is able to enable a fit in the presence of a data-model mismatch with a physically consistent interpretation with respect to the correct fit.

\begin{figure}[hbt!]
     \centering
     \includegraphics[clip,width=0.5\textwidth]{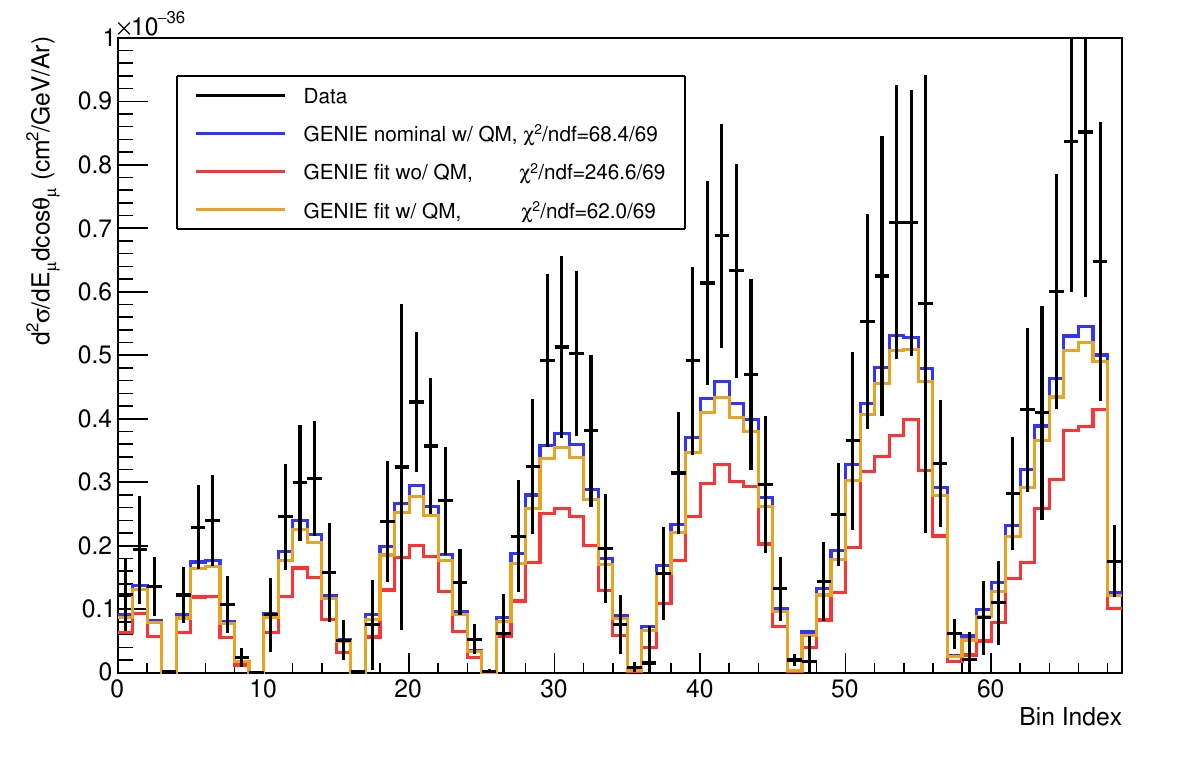}
     \put(-100,158){\normalsize{MicroBooNE Data}}
     \put(-200,105){\footnotesize{No $A_{C}$ Matrix}}
     \caption{Comparison of the measured MicroBooNE $d^{2}\sigma/dE_{\mu}d\cos\theta_{\mu}$ cross section data (black) with uncertainties enlarged through QM to the nominal \texttt{GENIE} model prediction without applying the regularization matrix (blue). Also plotted are fit results of the \texttt{GENIE} model without applying the regularization matrix tuned to data with (green) and without (red) using QM. In each case where QM is used in the fit it is also included in the reported $\chi^{2}$ calculation, and when QM is omitted from the fit it is also omitted from the $\chi^{2}$ calculation. The x-axis represents the bin index.}
    \label{fig:MicroBooNE_EMuCosThetaMu_noac_QM}
\end{figure}

Furthermore, QM increases the overall covariance so that the fit model prediction is able to describe the data within uncertainties, as shown in \autoref{fig:MicroBooNE_EMuCosThetaMu_noac_QM}.  This is especially useful when fitting to multiple data sets where one or more exhibit PPP symptoms, as this conservative approach allows for measurements that cannot be properly compared to a model prediction to be included in a joint fit without overwhelming the fit with PPP issues.

\section{Discussions}\label{sec:recommend}


In \autoref{sec:mismatch}, we explored several scenarios that can lead to data-model mismatches and provided recommendations for avoiding such discrepancies. In \autoref{sec:studies}, we demonstrated these scenarios through a series of case studies. However, it is impractical to enumerate every possible way data-model mismatches might occur, particularly when miscommunication arises between model tuning teams and analyzers reporting cross sections. To address this challenge, it is highly beneficial for analyzers reporting cross sections to also perform model fitting exercises, ensuring greater consistency and alignment between the reported data and model predictions.

Currently, it is common in cross section papers to report the comparison between the extracted cross sections and central-value model predictions from theoretical models, such as event generators. Test statistics, such as $\chi^{2}$, used to measure the discrepancies between data and central-value model predictions are often reported and used to draw qualitative conclusions regarding the model predictions. Building on this existing practice, we propose extending it to include uncertainties on the model predictions. This can be achieved using the \texttt{NUISANCE} package~\cite{Stowell:2016jfr}, which is commonly utilized for releasing extracted cross sections. Incorporating model prediction uncertainties not only enhances the ability to identify PPP-like issues but also facilitates a more thorough evaluation of the models’ accuracy and validity. For this exercise, we recommend to adopt the method outlined in \autoref{sec:methodology}, which introduces global and differential goodness-of-fit metrics. These metrics provide a systematic framework for assessing data-model agreement, offering additional insights into potential mismatches and improving the overall robustness of the analysis. Additionally, when fitting models to data, we recommend using as an aid the covariance matrix fit approach with physical boundaries enforced, as described in Sec.~\ref{sec:cov_matrix}. This allows for computationally efficient model fitting that removes the need to limit the number of model parameters included in the fit.

Looking forward, avoiding mismatches between model predictions and reported cross sections is essential to ensure accurate analyses. However, existing results have already been affected by such mismatches, including missing $A_C$ matrices and discrepancies between real and nominal neutrino flux. To address these challenges, we propose the Quantile Mapping (QM) technique, which introduces additional uncertainties to the data to mitigate inconsistencies between the data and model predictions. The QM approach conservatively reduces the weight of affected data and ensures that potential mismatches do not unduly bias model fits and is particularly useful in combining multiple datasets. By allowing the inclusion of datasets with suspected mismatches while appropriately reducing their influence, QM facilitates robust joint analyses and meaningful comparisons even when inconsistencies are present.

\section{Summary}\label{sec:summary}

This paper addresses the critical challenge of improving neutrino-nuclei interaction modeling for high-precision neutrino oscillation experiments by investigating Peelle’s Pertinent Puzzle, a persistent issue stemming from inconsistencies between experimental cross section data and theoretical models. Through case studies, the paper identifies key sources of these inconsistencies, including the omission of the regularization matrix $A_C$ in data unfolding, discrepancies between real and nominal neutrino fluxes, and limitations in model coverage. These issues can lead to significant data-model mismatches, hindering the reliability of model predictions.

To mitigate these challenges, we recommend that analysts avoid the publication of measurements that necessitate known improper treatments such as the omission of the regularization matrix and the use of real-flux measurements without supporting flux details. We also recommend incorporating strategies such as model fitting exercises into cross section publications and leveraging the Quantile Mapping technique, which introduces additional uncertainties to account for inconsistencies while reducing their impact on model fits. The \texttt{NUISANCE} framework is highlighted as a valuable tool for performing systematic fits and identifying PPP-related challenges. Building on top of this, the paper introduces a new model fitting technique that minimizes a test statistic constructed from the data, model prediction, and their covariances. Describing model variations through a covariance matrix allows the simultaneous fitting of an arbitrarily large number of model parameters. By ensuring consistent treatment of data and model predictions, these methods facilitate robust joint analyses and meaningful comparisons across datasets. The findings emphasize the importance of addressing typical data-model inconsistencies and provide insights for advancing the accuracy and reliability of neutrino-nuclei interaction models, critical for the success of high-precision neutrino experiments.


\begin{acknowledgments}
We would like to acknowledge the support and valuable discussions provided during the \href{https://indico.fnal.gov/event/64511/}{Generator Studies Workshop} at the Pittsburgh Particle Physics, Astrophysics, and Cosmology Center (PittPACC). The workshop served as a highly collaborative platform that facilitated insightful exchanges and inspired advancements in this work. We deeply appreciate PittPACC’s efforts in fostering a vibrant scientific community and enabling critical dialogues that contribute significantly to progress in the field.
\end{acknowledgments}
\vspace{-3mm}

\appendix

\section{Parameter Fit Values}

\begin{table*}[!htb]
    \centering
    \caption{\texttt{GENIE} parameter physical values after fits to the CC inclusive \texttt{MicroBooNE} measurements~\cite{PhysRevLett.133.041801,MicroBooNE:2023foc} and the CC pionless \texttt{T2K} measurement~\cite{T2K:2020jav}.}
    \begin{tabular}{|l|c|c|c|c|}
    \hline
    Fit                                                                                               & $M_{A}^{QE}$\,(GeV) & RPA             & CCMEC Norm       & CCMEC Shape     \\
    \hline
    Nominal                                                                                           & $1.1  \pm 0.1$      & $0.85 \pm 0.15$ &  $1.66 \pm 0.5$  &  $1^{+0}_{-1}$   \\
    \texttt{MicroBooNE} $d\sigma/dE_{\mu}$                                                            & $1.18 \pm 0.08$     & $1.08 \pm 0.36$ &  $1.37 \pm 0.43$ & $-0.13 \pm 1.04$ \\
    \texttt{MicroBooNE} $d\sigma/dE_{\mu}$ w/ no $A_{C}$                                              & $1.15 \pm 0.08$     & $1.21 \pm 0.34$ &  $1.29 \pm 0.41$ &  $0.05 \pm 1.00$ \\
    \texttt{MicroBooNE} $d^{2}\sigma/dE_{\mu}d\cos\theta_{\mu}$                                       & $1.26 \pm 0.07$     & $0.72 \pm 0.25$ &  $1.67 \pm 0.38$ &  $0.09 \pm 0.85$ \\
    \texttt{MicroBooNE} $d^{2}\sigma/dE_{\mu}d\cos\theta_{\mu}$ w/ no $A_{C}$                         & $1.14 \pm 0.04$     & $0.12 \pm 0.13$ & $-0.85 \pm 0.23$ &  $3.15 \pm 0.56$ \\
    \texttt{MicroBooNE} $d^{2}\sigma(E_{\nu})/dP_{\mu}d\cos\theta_{\mu}$                              & $1.36 \pm 0.05$     & $0.14 \pm 0.18$ &  $1.68 \pm 0.34$ &  $0.05 \pm 0.66$ \\
    \texttt{MicroBooNE} $d^{2}\sigma(E_{\nu})/dP_{\mu}d\cos\theta_{\mu}$ w/ no $A_{C}$                & $0.66 \pm 0.02$     & $1.05 \pm 0.07$ &  $0.25 \pm 0.15$ &  $1.29 \pm 0.52$ \\
    \hline
    \texttt{MicroBooNE} $d^{2}\sigma(E_{\nu})/dP_{\mu}d\cos\theta_{\mu}$ w/ alt flux                  & $1.07 \pm 0.05$     & $0.69 \pm 0.18$ &  $1.10 \pm 0.28$ &  $1.33 \pm 0.82$ \\
    \hline
    \texttt{MicroBooNE} $d^{2}\sigma(E_{\nu})/dP_{\mu}d\cos\theta_{\mu}$ w/ QM, no $A_{C}$            & $0.99 \pm 0.09$ & $0.76 \pm 0.35$ &  $0.61 \pm 0.44$ & $1.51 \pm 0.88$ \\
    \texttt{MicroBooNE} $d^{2}\sigma(E_{\nu})/dP_{\mu}d\cos\theta_{\mu}$ w/ QM, reordered, no $A_{C}$ & $1.02 \pm 0.09$ & $0.65 \pm 0.34$ &  $0.98 \pm 0.39$ & $1.58 \pm 0.88$ \\
    \hline
    \texttt{T2K} $d^{2}\sigma/dP_{\mu}d\cos\theta_{\mu}$                                              & $1.12 \pm 0.10$ & $0.75 \pm 0.26$ &  $1.37 \pm 0.43$ & $0.48 \pm 0.37$ \\
    \texttt{T2K} $d^{2}\sigma/dP_{\mu}d\cos\theta_{\mu}$ w/ argon model                               & $0.53 \pm 0.14$ & $1.32 \pm 0.23$ &  $1.37 \pm 0.43$ & $0.94 \pm 0.71$ \\
    \hline
    \end{tabular}
    \label{table:genie_pars}
\end{table*}

\begin{table*}[!htb]
    \centering
    \caption{\texttt{NEUT} parameter physical values after fits to the CC inclusive \texttt{MicroBooNE} measurement~\cite{PhysRevLett.133.041801} and the CC pionless \texttt{T2K} measurement~\cite{T2K:2020jav}. Due to the way the fits of the NEUT model were implemented, it is not feasible to accurately quote parameter uncertainties.}
    \begin{tabular}{|l|c|c|c|c|c|c|c|c|c|}
    \hline
                                                                         & $M_{A}^{QE}$\,(GeV) & Optical Potential & High-$Q^{2}$ Norm & 2p2h Norm & $M_{A}^{RES}$\,(GeV) & $C^A_{5}$ \\
    \hline
    Nominal                                                              & $1.21$              & $0.00$            & $1.00$            & $1.00$    & $0.95$               & $1.01$ \\
    \texttt{T2K} $d^{2}\sigma/dP_{\mu}d\cos\theta_{\mu}$                 & $0.56$              & $1.00$            & $1.96$            & $0.13$    & $1.08$               & $1.10$ \\
    \texttt{T2K} $d^{2}\sigma/dP_{\mu}d\cos\theta_{\mu}$ expanded        & $0.70$              & $0.90$            & $1.51$            & $0.54$    & $0.44$               & $2.00$ \\
    \texttt{T2K} $d^{2}\sigma/dP_{\mu}d\cos\theta_{\mu}$ w/ Nieves       & $0.79$              & $0.00$            & $1.19$            & $0.57$    & $0.86$               & $1.30$ \\
    \texttt{MicroBooNE} $d^{2}\sigma(E_{\nu})/dP_{\mu}d\cos\theta_{\mu}$ & $1.45$              & $0.00$            & $1.00$            & $1.00$    & $0.95$               & $1.00$ \\
    \hline
    \end{tabular}
    \label{table:neut_pars}
\end{table*}

A selection of parameter fit values for each of the model fit attempts detailed above are presented in Table~\ref{table:genie_pars} and Table~\ref{table:neut_pars} along with their uncorrelated uncertainties. Given the number of parameters fit in some instances, it is not practical to present all fit parameter, therefore only a subset are shown. Specifically, in fits of \texttt{GENIE} predictions, only the four \texttt{GENIE} parameters used in the tuning of the \texttt{MicroBooNE model}~\cite{MicroBooNE:2021ccs} are shown. Note that fits performed by minimizing a $\chi^{2}$ test statistic computed with the covariance matrix do not inherently constrain parameter values within their physical bounds. As a result, some post-fit parameter values are non-physical. This issue is only significantly observed in cases where improper treatments are artificially added, and could be avoided through a more sophisticated test statistic minimization that incorporated parameter boundaries, as discussed in Sec.~\ref{sec:cov_matrix}.

Fit parameters are shown only to give further insight into the degree of PPP symptoms observed, not to promote a particular model tune for any physics goal. For example, in some cases where inconsistencies are introduced non-physical parameter fit values are observed. These fit parameters are intended as illustrative rather than prescriptive for multiple reasons, including their incomplete listing, non-physical fitted parameter values, the limitation of only presenting uncorrelated uncertainties rather than the full covariance between parameters, and the fact that correlations between cross section models used to unfold some measurements and fitted models were not considered in fitting.

\section{Quantile Mapping Details}

QM considers a given random variable $X$ with probability distribution function (PDF) $f_{X}$ and cumulative distribution function (CDF) $g_{X}$.  The goal of QM is to employ a mapping that transforms the PDF $f_{X}$ into a chosen target distribution $f_{Y}$ with CDF $g_{Y}$.  This is achieved by ensuring that $g_{X}$ maps to $g_{Y}$, which guarantees agreement between the distributions $f_{X}$ and $f_{Y}$ as well.  Consider the mapping $X \rightarrow \widetilde{X}$:
\begin{equation}
    \widetilde{X} = g^{-1}_{Y} (g_{X}(X)).
\end{equation}
Since $g_{X}(X)$ and therefore $g_{Y}(\widetilde{X})$ are uniformly distributed on $[0,1]$, $\widetilde{X}$ has the same CDF $g_{Y}$ as $Y$.  Therefore, for any well-behaved $f_{X}$, a mapping can be constructed to yield the desired PDF $f_{Y}$ so long as $g^{-1}_{Y}$ and $g_{X}$ are known.

In this application, the given random variable of interest is the uncertainty-weighted difference between measurement and prediction in the diagonalized basis of uncorrelated bins.  This allows each $\chi^{2}_{i}$ term to be taken as a throw from the PDF of X:
\begin{equation}
    f_{X}(x) = k(x)/n,
\end{equation}
\begin{equation}
    g_{X}(x) = K(x)/n,
\end{equation}
where $(x)$ counts the number of bins with a $\chi^{2}_{i}$ value of $x$, $K(x)$ counts the number of bins with a $\chi^{2}_{i}$ value less than or equal to $x$, and the domain $\mathbb{R} \geq 0$ includes all  possible $\chi^{2}_{i}$ values.  Given the finite number of bins, $k(x)$ will necessarily be 0 for most values of $x$.  We are interested in constructing a covariance $\widetilde{\Sigma}_{D}$ that is able to describe the observed discrepancy $D$ between measurement and prediction, therefore individual $\chi^{2}_{i}$ terms under $\widetilde{\Sigma}_{D}$ should follow a $\chi^{2}$ distribution with one DoF, and we can set:
\begin{equation}
    f_{Y}(x) = \chi^{2}(x,1).
\end{equation}
Rather than computing $g^{-1}_{Y}$ from $f_{Y}$, it is easier to determine its form by inspecting the mapping performed by $g_{Y}$.  Specifically, note that $g_{Y}$ maps the $\chi^{2}(x,1)$ distribution to $p$, where $p$ represents the $p$-value of sampling a $\chi^{2}$ value of $x$ or smaller.  Conversely, $g^{-1}_{Y}$ maps a $p$-value to its corresponding $\chi^{2}$ value.  For one DoF, the $\chi^{2}$ distribution is generated from the square of a single normally distributed variable, and thus the relation between $p$-value and $\chi^{2}$ value can be computed from the error function $Erf$:
\begin{equation}
    g^{-1}_{Y}(p) = 2\left( Erf^{-1}(p) \right)^{2}.
\end{equation}

It is now possible to determine the required mapping $X \rightarrow \widetilde{X}$ so that $\widetilde{X}$ will be $\chi^{2}$ distributed with one DoF.  To achieve this transformation the covariance $\Sigma_{D,ii}$ of each bin is scaled by $\chi^{2}_{i}/g^{-1}_{Y}(g(\chi^{2}_{i}))$ while the measured and predicted cross sections, and therefore $D_{i}$, are left unchanged.  Since the initial issue was an insufficient covariance to describe the difference between measurement and prediction, this will naturally lead to an overall enlargement of the covariance describing the comparison, which can be thought of as an enlargement of the uncertainty in the data measurement.  Furthermore, one can deviate from a strict implementation of QM by declining to reduce the uncertainty of any bins whenever QM calls for such a reduction.  This makes the application of QM a conservative strategy for mitigating tensions between data and model.

\section{Uncertainty Reordering}

In practice, QM proves to be effective at reducing the impact of an improper treatment and can prevent a PPP normalization issue, but does not always lead to a fit result that significantly improves on the pre-fit model prediction.  One way to understand this outcome is to realize that QM merely re-scales each $\chi^{2}_{i}$ value without changing their ordering from smallest to largest.  As a result, bins that previously exhibited extreme tension from an improper treatment will still exhibit large tension after QM but to a reduced degree. This tension may still be significant enough to dis-incentivize a fit from addressing disagreements in other DoF.  The result is that a fit attempt, even after performing QM, may be biased by non-physical discrepancies resulting from an improper treatment rather than focusing on real physics information present in a measurement.

To resolve this issue, an analyst may apply their own intuition to give preference to the fitting of certain DoFs over others.  In general this is a difficult task as it requires deciding which discrepancies in the diagonalized space are most physically relevant. However, it may be reasonable to increase the significance of a large eigenvalue DoF such as the normalization. Mechanically, this can be achieved by reducing the uncertainty for one or more DoF before applying QM. Then QM will ensure that the overall covariance is able to describe the data-model discrepancy, but will place any preferenced DoF in higher tension than otherwise, at the expense of other DoF.

\begin{figure}[hbt!]
     \centering
     \includegraphics[clip,width=0.5\textwidth]{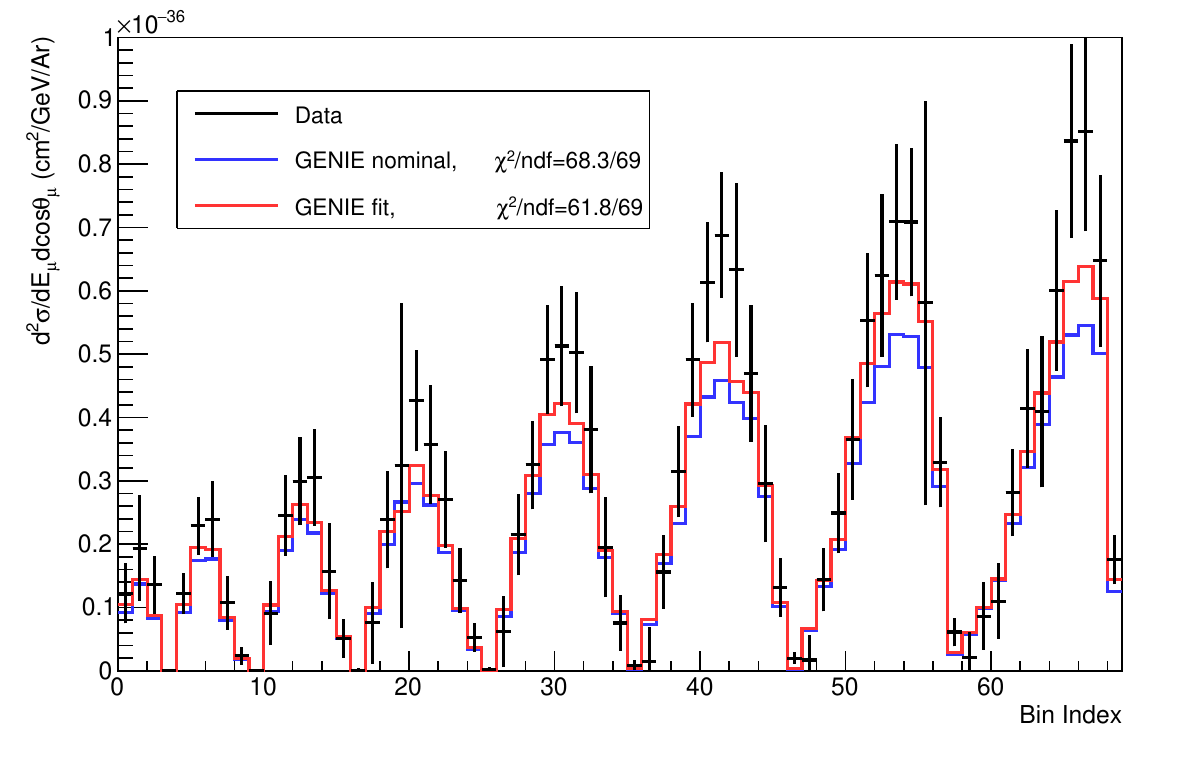}
     \put(-100, 158){\normalsize MicroBooNE Data}
     \put(-200,107){\footnotesize{No $A_{C}$ Matrix}}
     \caption{Comparison of the measured MicroBooNE $d^{2}\sigma/dE_{\mu}d\cos\theta_{\mu}$ cross section data (black) to the nominal model prediction without applying the regularization matrix (blue), and the fit result without applying the regularization matrix (red). The x-axis represents the bin index. The PPP-mitigation strategy combining uncertainty reordering and quantile mapping allows for a somewhat successful fit despite the mismatch between data and model.}
    \label{fig:MicroBooNE_EMuCosThetaMu_noac_QM_unc_reorder}
\end{figure}

We demonstrate an implementation of uncertainty reordering for the example of the previously mentioned inclusive $\nu_{\mu}$CC double differential cross section $d^{2}\sigma/dE_{\mu}d\cos\theta_{\mu}$ in the case where the regularization matrix is omitted from the model prediction. Previously, this omission was demonstrated to lead to a severe PPP issue in the overall normalization of the post-fit model. Furthermore, the use of QM was found to reduce this PPP issue, but visually the post-fit model did not significantly improve the comparison to the data compared to the nominal model prediction.  In this example the largest eigenvalue DoF (referred to here as normalization) is prioritized by halving its uncertainty, in effect increasing the corresponding $\chi^{2}_{i}$ by a factor of four.  Quantile mapping is then applied, which increases the covariance in many DoF including the normalization, but keeps the newly established relative order among DoF. As a result, the fit gives higher significance to preserving a good normalization agreement, shown in Fig.~\ref{fig:MicroBooNE_EMuCosThetaMu_noac_QM_unc_reorder}. A better visual agreement between post-fit model and data is observed in this instance, approaching the visual fit quality observed when the regularization matrix is included. In this example the use of uncertainty reordering was straightforward as only the normalization was addressed, however in principle this method may be more difficult to successfully apply based on the complexity of the measurement and the nature of the underlying improper treatment between the data and model.


\begin{thebibliography}{54}%
\makeatletter
\providecommand \@ifxundefined [1]{%
 \@ifx{#1\undefined}
}%
\providecommand \@ifnum [1]{%
 \ifnum #1\expandafter \@firstoftwo
 \else \expandafter \@secondoftwo
 \fi
}%
\providecommand \@ifx [1]{%
 \ifx #1\expandafter \@firstoftwo
 \else \expandafter \@secondoftwo
 \fi
}%
\providecommand \natexlab [1]{#1}%
\providecommand \enquote  [1]{``#1''}%
\providecommand \bibnamefont  [1]{#1}%
\providecommand \bibfnamefont [1]{#1}%
\providecommand \citenamefont [1]{#1}%
\providecommand \href@noop [0]{\@secondoftwo}%
\providecommand \href [0]{\begingroup \@sanitize@url \@href}%
\providecommand \@href[1]{\@@startlink{#1}\@@href}%
\providecommand \@@href[1]{\endgroup#1\@@endlink}%
\providecommand \@sanitize@url [0]{\catcode `\\12\catcode `\$12\catcode `\&12\catcode `\#12\catcode `\^12\catcode `\_12\catcode `\%12\relax}%
\providecommand \@@startlink[1]{}%
\providecommand \@@endlink[0]{}%
\providecommand \url  [0]{\begingroup\@sanitize@url \@url }%
\providecommand \@url [1]{\endgroup\@href {#1}{\urlprefix }}%
\providecommand \urlprefix  [0]{URL }%
\providecommand \Eprint [0]{\href }%
\providecommand \doibase [0]{https://doi.org/}%
\providecommand \selectlanguage [0]{\@gobble}%
\providecommand \bibinfo  [0]{\@secondoftwo}%
\providecommand \bibfield  [0]{\@secondoftwo}%
\providecommand \translation [1]{[#1]}%
\providecommand \BibitemOpen [0]{}%
\providecommand \bibitemStop [0]{}%
\providecommand \bibitemNoStop [0]{.\EOS\space}%
\providecommand \EOS [0]{\spacefactor3000\relax}%
\providecommand \BibitemShut  [1]{\csname bibitem#1\endcsname}%
\let\auto@bib@innerbib\@empty
\bibitem [{\citenamefont {Abe}\ \emph {et~al.}(2023)\citenamefont {Abe} \emph {et~al.}}]{T2K:2023smv}%
  \BibitemOpen
  \bibfield  {author} {\bibinfo {author} {\bibfnamefont {K.}~\bibnamefont {Abe}} \emph {et~al.} (\bibinfo {collaboration} {T2K Collaboration}),\ }\bibfield  {title} {\bibinfo {title} {{Measurements of neutrino oscillation parameters from the T2K experiment using $3.6\times 10^{21}$ protons on target}},\ }\href {https://doi.org/10.1140/epjc/s10052-023-11819-x} {\bibfield  {journal} {\bibinfo  {journal} {Eur. Phys. J. C}\ }\textbf {\bibinfo {volume} {83}},\ \bibinfo {pages} {782} (\bibinfo {year} {2023})}\BibitemShut {NoStop}%
\bibitem [{\citenamefont {Acero}\ \emph {et~al.}(2024)\citenamefont {Acero} \emph {et~al.}}]{NOvA:2023iam}%
  \BibitemOpen
  \bibfield  {author} {\bibinfo {author} {\bibfnamefont {M.~A.}\ \bibnamefont {Acero}} \emph {et~al.} (\bibinfo {collaboration} {NOvA}),\ }\bibfield  {title} {\bibinfo {title} {{Expanding neutrino oscillation parameter measurements in NOvA using a Bayesian approach}},\ }\href {https://doi.org/10.1103/PhysRevD.110.012005} {\bibfield  {journal} {\bibinfo  {journal} {Phys. Rev. D}\ }\textbf {\bibinfo {volume} {110}},\ \bibinfo {pages} {012005} (\bibinfo {year} {2024})}\BibitemShut {NoStop}%
\bibitem [{\citenamefont {Abi}\ \emph {et~al.}(2020)\citenamefont {Abi} \emph {et~al.}}]{DUNE:2020jqi}%
  \BibitemOpen
  \bibfield  {author} {\bibinfo {author} {\bibfnamefont {B.}~\bibnamefont {Abi}} \emph {et~al.} (\bibinfo {collaboration} {DUNE}),\ }\bibfield  {title} {\bibinfo {title} {{Long-baseline neutrino oscillation physics potential of the DUNE experiment}},\ }\href {https://doi.org/10.1140/epjc/s10052-020-08456-z} {\bibfield  {journal} {\bibinfo  {journal} {Eur. Phys. J. C}\ }\textbf {\bibinfo {volume} {80}},\ \bibinfo {pages} {978} (\bibinfo {year} {2020})}\BibitemShut {NoStop}%
\bibitem [{\citenamefont {Abe}\ \emph {et~al.}(2018{\natexlab{a}})\citenamefont {Abe} \emph {et~al.}}]{Hyper-Kamiokande:2018ofw}%
  \BibitemOpen
  \bibfield  {author} {\bibinfo {author} {\bibfnamefont {K.}~\bibnamefont {Abe}} \emph {et~al.} (\bibinfo {collaboration} {Hyper-Kamiokande}),\ }\bibfield  {title} {\bibinfo {title} {{Hyper-Kamiokande Design Report}},\ }\Eprint {https://arxiv.org/abs/1805.04163} {arXiv:1805.04163 [physics.ins-det]}  (\bibinfo {year} {2018}{\natexlab{a}})\BibitemShut {NoStop}%
\bibitem [{\citenamefont {Abe}\ \emph {et~al.}(2020{\natexlab{a}})\citenamefont {Abe} \emph {et~al.}}]{Abe:2019vii}%
  \BibitemOpen
  \bibfield  {author} {\bibinfo {author} {\bibfnamefont {K.}~\bibnamefont {Abe}} \emph {et~al.} (\bibinfo {collaboration} {T2K}),\ }\bibfield  {title} {\bibinfo {title} {{Constraint on the matter-antimatter symmetry-violating phase in neutrino oscillations}},\ }\href {https://doi.org/10.1038/s41586-020-2177-0} {\bibfield  {journal} {\bibinfo  {journal} {Nature}\ }\textbf {\bibinfo {volume} {580}},\ \bibinfo {pages} {339} (\bibinfo {year} {2020}{\natexlab{a}})}\BibitemShut {NoStop}%
\bibitem [{\citenamefont {Qian}\ and\ \citenamefont {Vogel}(2015)}]{Qian:2015waa}%
  \BibitemOpen
  \bibfield  {author} {\bibinfo {author} {\bibfnamefont {X.}~\bibnamefont {Qian}}\ and\ \bibinfo {author} {\bibfnamefont {P.}~\bibnamefont {Vogel}},\ }\bibfield  {title} {\bibinfo {title} {{Neutrino Mass Hierarchy}},\ }\href {https://doi.org/10.1016/j.ppnp.2015.05.002} {\bibfield  {journal} {\bibinfo  {journal} {Prog. Part. Nucl. Phys.}\ }\textbf {\bibinfo {volume} {83}},\ \bibinfo {pages} {1} (\bibinfo {year} {2015})}\BibitemShut {NoStop}%
\bibitem [{\citenamefont {Abazajian}\ \emph {et~al.}(2012)\citenamefont {Abazajian} \emph {et~al.}}]{Abazajian:2012ys}%
  \BibitemOpen
  \bibfield  {author} {\bibinfo {author} {\bibfnamefont {K.~N.}\ \bibnamefont {Abazajian}} \emph {et~al.},\ }\bibfield  {title} {\bibinfo {title} {{Light Sterile Neutrinos: A White Paper}},\ }\bibfield  {journal} {\bibinfo  {journal} {arXiv e-prints}\ }\href {https://doi.org/10.48550/arXiv.1204.5379} {10.48550/arXiv.1204.5379} (\bibinfo {year} {2012})\BibitemShut {NoStop}%
\bibitem [{\citenamefont {Di~Lodovico}\ \emph {et~al.}(2023)\citenamefont {Di~Lodovico}, \citenamefont {Patterson}, \citenamefont {Shiozawa},\ and\ \citenamefont {Worcester}}]{DiLodovico:2023jgr}%
  \BibitemOpen
  \bibfield  {author} {\bibinfo {author} {\bibfnamefont {F.}~\bibnamefont {Di~Lodovico}}, \bibinfo {author} {\bibfnamefont {R.~B.}\ \bibnamefont {Patterson}}, \bibinfo {author} {\bibfnamefont {M.}~\bibnamefont {Shiozawa}},\ and\ \bibinfo {author} {\bibfnamefont {E.}~\bibnamefont {Worcester}},\ }\bibfield  {title} {\bibinfo {title} {{Experimental Considerations in Long-Baseline Neutrino Oscillation Measurements}},\ }\href {https://doi.org/10.1146/annurev-nucl-102020-101615} {\bibfield  {journal} {\bibinfo  {journal} {Ann. Rev. Nucl. Part. Sci.}\ }\textbf {\bibinfo {volume} {73}},\ \bibinfo {pages} {69} (\bibinfo {year} {2023})}\BibitemShut {NoStop}%
\bibitem [{\citenamefont {Workman}\ \emph {et~al.}(2022)\citenamefont {Workman} \emph {et~al.}}]{Workman:2022ynf}%
  \BibitemOpen
  \bibfield  {author} {\bibinfo {author} {\bibfnamefont {R.~L.}\ \bibnamefont {Workman}} \emph {et~al.} (\bibinfo {collaboration} {Particle Data Group}),\ }\bibfield  {title} {\bibinfo {title} {{Review of Particle Physics}},\ }\href {https://doi.org/10.1093/ptep/ptac097} {\bibfield  {journal} {\bibinfo  {journal} {PTEP}\ }\textbf {\bibinfo {volume} {2022}},\ \bibinfo {pages} {083C01} (\bibinfo {year} {2022})}\BibitemShut {NoStop}%
\bibitem [{\citenamefont {Formaggio}\ and\ \citenamefont {Zeller}(2012)}]{Formaggio:2012cpf}%
  \BibitemOpen
  \bibfield  {author} {\bibinfo {author} {\bibfnamefont {J.~A.}\ \bibnamefont {Formaggio}}\ and\ \bibinfo {author} {\bibfnamefont {G.~P.}\ \bibnamefont {Zeller}},\ }\bibfield  {title} {\bibinfo {title} {{From eV to EeV: Neutrino Cross Sections Across Energy Scales}},\ }\href {https://doi.org/10.1103/RevModPhys.84.1307} {\bibfield  {journal} {\bibinfo  {journal} {Rev. Mod. Phys.}\ }\textbf {\bibinfo {volume} {84}},\ \bibinfo {pages} {1307} (\bibinfo {year} {2012})}\BibitemShut {NoStop}%
\bibitem [{\citenamefont {Mosel}(2016)}]{Mosel:2016cwa}%
  \BibitemOpen
  \bibfield  {author} {\bibinfo {author} {\bibfnamefont {U.}~\bibnamefont {Mosel}},\ }\bibfield  {title} {\bibinfo {title} {{Neutrino Interactions with Nucleons and Nuclei: Importance for Long-Baseline Experiments}},\ }\href {https://doi.org/10.1146/annurev-nucl-102115-044720} {\bibfield  {journal} {\bibinfo  {journal} {Ann. Rev. Nucl. Part. Sci.}\ }\textbf {\bibinfo {volume} {66}},\ \bibinfo {pages} {171} (\bibinfo {year} {2016})}\BibitemShut {NoStop}%
\bibitem [{\citenamefont {Gross}\ \emph {et~al.}(2023)\citenamefont {Gross} \emph {et~al.}}]{Gross:2022hyw}%
  \BibitemOpen
  \bibfield  {author} {\bibinfo {author} {\bibfnamefont {F.}~\bibnamefont {Gross}} \emph {et~al.},\ }\bibfield  {title} {\bibinfo {title} {{50 Years of Quantum Chromodynamics}},\ }\href {https://doi.org/doi.org/10.1140/epjc/s10052-023-11949-2} {\bibfield  {journal} {\bibinfo  {journal} {Eur. Phys. J. C}\ }\textbf {\bibinfo {volume} {83}} (\bibinfo {year} {2023})}\BibitemShut {NoStop}%
\bibitem [{\citenamefont {Alvarez-Ruso}\ \emph {et~al.}(2021)\citenamefont {Alvarez-Ruso} \emph {et~al.}}]{GENIE:2021npt}%
  \BibitemOpen
  \bibfield  {author} {\bibinfo {author} {\bibfnamefont {L.}~\bibnamefont {Alvarez-Ruso}} \emph {et~al.} (\bibinfo {collaboration} {GENIE}),\ }\bibfield  {title} {\bibinfo {title} {{Recent highlights from GENIE v3}},\ }\href {https://doi.org/10.1140/epjs/s11734-021-00295-7} {\bibfield  {journal} {\bibinfo  {journal} {Eur. Phys. J. ST}\ }\textbf {\bibinfo {volume} {230}},\ \bibinfo {pages} {4449} (\bibinfo {year} {2021})}\BibitemShut {NoStop}%
\bibitem [{\citenamefont {Hayato}\ and\ \citenamefont {Pickering}(2021)}]{Hayato:2021heg}%
  \BibitemOpen
  \bibfield  {author} {\bibinfo {author} {\bibfnamefont {Y.}~\bibnamefont {Hayato}}\ and\ \bibinfo {author} {\bibfnamefont {L.}~\bibnamefont {Pickering}},\ }\bibfield  {title} {\bibinfo {title} {{The NEUT neutrino interaction simulation program library}},\ }\href {https://doi.org/10.1140/epjs/s11734-021-00287-7} {\bibfield  {journal} {\bibinfo  {journal} {Eur. Phys. J. ST}\ }\textbf {\bibinfo {volume} {230}},\ \bibinfo {pages} {4469} (\bibinfo {year} {2021})}\BibitemShut {NoStop}%
\bibitem [{\citenamefont {Golan}\ \emph {et~al.}(2012)\citenamefont {Golan}, \citenamefont {Sobczyk},\ and\ \citenamefont {Zmuda}}]{Golan:2012rfa}%
  \BibitemOpen
  \bibfield  {author} {\bibinfo {author} {\bibfnamefont {T.}~\bibnamefont {Golan}}, \bibinfo {author} {\bibfnamefont {J.~T.}\ \bibnamefont {Sobczyk}},\ and\ \bibinfo {author} {\bibfnamefont {J.}~\bibnamefont {Zmuda}},\ }\bibfield  {title} {\bibinfo {title} {{NuWro: the Wroclaw Monte Carlo Generator of Neutrino Interactions}},\ }\href {https://doi.org/10.1016/j.nuclphysbps.2012.09.136} {\bibfield  {journal} {\bibinfo  {journal} {Nucl. Phys. B Proc. Suppl.}\ }\textbf {\bibinfo {volume} {229-232}},\ \bibinfo {pages} {499} (\bibinfo {year} {2012})}\BibitemShut {NoStop}%
\bibitem [{\citenamefont {Buss}\ \emph {et~al.}(2012)\citenamefont {Buss}, \citenamefont {Gaitanos}, \citenamefont {Gallmeister}, \citenamefont {van Hees}, \citenamefont {Kaskulov}, \citenamefont {Lalakulich}, \citenamefont {Larionov}, \citenamefont {Leitner}, \citenamefont {Weil},\ and\ \citenamefont {Mosel}}]{Buss:2011mx}%
  \BibitemOpen
  \bibfield  {author} {\bibinfo {author} {\bibfnamefont {O.}~\bibnamefont {Buss}}, \bibinfo {author} {\bibfnamefont {T.}~\bibnamefont {Gaitanos}}, \bibinfo {author} {\bibfnamefont {K.}~\bibnamefont {Gallmeister}}, \bibinfo {author} {\bibfnamefont {H.}~\bibnamefont {van Hees}}, \bibinfo {author} {\bibfnamefont {M.}~\bibnamefont {Kaskulov}}, \bibinfo {author} {\bibfnamefont {O.}~\bibnamefont {Lalakulich}}, \bibinfo {author} {\bibfnamefont {A.~B.}\ \bibnamefont {Larionov}}, \bibinfo {author} {\bibfnamefont {T.}~\bibnamefont {Leitner}}, \bibinfo {author} {\bibfnamefont {J.}~\bibnamefont {Weil}},\ and\ \bibinfo {author} {\bibfnamefont {U.}~\bibnamefont {Mosel}},\ }\bibfield  {title} {\bibinfo {title} {{Transport-theoretical Description of Nuclear Reactions}},\ }\href {https://doi.org/10.1016/j.physrep.2011.12.001} {\bibfield  {journal} {\bibinfo  {journal} {Phys. Rept.}\ }\textbf {\bibinfo {volume} {512}},\ \bibinfo {pages} {1} (\bibinfo {year} {2012})}\BibitemShut {NoStop}%
\bibitem [{\citenamefont {Isaacson}\ \emph {et~al.}(2023)\citenamefont {Isaacson}, \citenamefont {Jay}, \citenamefont {Lovato}, \citenamefont {Machado},\ and\ \citenamefont {Rocco}}]{PhysRevD.107.033007}%
  \BibitemOpen
  \bibfield  {author} {\bibinfo {author} {\bibfnamefont {J.}~\bibnamefont {Isaacson}}, \bibinfo {author} {\bibfnamefont {W.~I.}\ \bibnamefont {Jay}}, \bibinfo {author} {\bibfnamefont {A.}~\bibnamefont {Lovato}}, \bibinfo {author} {\bibfnamefont {P.~A.~N.}\ \bibnamefont {Machado}},\ and\ \bibinfo {author} {\bibfnamefont {N.}~\bibnamefont {Rocco}},\ }\bibfield  {title} {\bibinfo {title} {Introducing a novel event generator for electron-nucleus and neutrino-nucleus scattering},\ }\href {https://doi.org/10.1103/PhysRevD.107.033007} {\bibfield  {journal} {\bibinfo  {journal} {Phys. Rev. D}\ }\textbf {\bibinfo {volume} {107}},\ \bibinfo {pages} {033007} (\bibinfo {year} {2023})}\BibitemShut {NoStop}%
\bibitem [{\citenamefont {Gallagher}(2006)}]{GALLAGHER2006229}%
  \BibitemOpen
  \bibfield  {author} {\bibinfo {author} {\bibfnamefont {H.}~\bibnamefont {Gallagher}},\ }\bibfield  {title} {\bibinfo {title} {Event generator tuning in the resonance region for the minos experiment},\ }\href {https://doi.org/https://doi.org/10.1016/j.nuclphysbps.2006.08.041} {\bibfield  {journal} {\bibinfo  {journal} {Nuclear Physics B - Proceedings Supplements}\ }\textbf {\bibinfo {volume} {159}},\ \bibinfo {pages} {229} (\bibinfo {year} {2006})},\ \bibinfo {note} {proceedings of the 4th International Workshop on Neutrino-Nucleus Interactions in the Few-GeV Region}\BibitemShut {NoStop}%
\bibitem [{\citenamefont {Acero}\ \emph {et~al.}(2020)\citenamefont {Acero} \emph {et~al.}}]{NOvA:2020rbg}%
  \BibitemOpen
  \bibfield  {author} {\bibinfo {author} {\bibfnamefont {M.~A.}\ \bibnamefont {Acero}} \emph {et~al.} (\bibinfo {collaboration} {NOvA, R. Group}),\ }\bibfield  {title} {\bibinfo {title} {{Adjusting neutrino interaction models and evaluating uncertainties using NOvA near detector data}},\ }\href {https://doi.org/10.1140/epjc/s10052-020-08577-5} {\bibfield  {journal} {\bibinfo  {journal} {Eur. Phys. J. C}\ }\textbf {\bibinfo {volume} {80}},\ \bibinfo {pages} {1119} (\bibinfo {year} {2020})}\BibitemShut {NoStop}%
\bibitem [{\citenamefont {Stowell}\ \emph {et~al.}(2019)\citenamefont {Stowell} \emph {et~al.}}]{MINERvA:2019kfr}%
  \BibitemOpen
  \bibfield  {author} {\bibinfo {author} {\bibfnamefont {P.}~\bibnamefont {Stowell}} \emph {et~al.} (\bibinfo {collaboration} {MINERvA}),\ }\bibfield  {title} {\bibinfo {title} {{Tuning the GENIE Pion Production Model with MINER$\nu$A Data}},\ }\href {https://doi.org/10.1103/PhysRevD.100.072005} {\bibfield  {journal} {\bibinfo  {journal} {Phys. Rev. D}\ }\textbf {\bibinfo {volume} {100}},\ \bibinfo {pages} {072005} (\bibinfo {year} {2019})}\BibitemShut {NoStop}%
\bibitem [{\citenamefont {Abratenko}\ \emph {et~al.}(2022{\natexlab{a}})\citenamefont {Abratenko} \emph {et~al.}}]{MicroBooNE:2021ccs}%
  \BibitemOpen
  \bibfield  {author} {\bibinfo {author} {\bibfnamefont {P.}~\bibnamefont {Abratenko}} \emph {et~al.} (\bibinfo {collaboration} {MicroBooNE}),\ }\bibfield  {title} {\bibinfo {title} {{New $CC0\pi$ GENIE model tune for MicroBooNE}},\ }\href {https://doi.org/10.1103/PhysRevD.105.072001} {\bibfield  {journal} {\bibinfo  {journal} {Phys. Rev. D}\ }\textbf {\bibinfo {volume} {105}},\ \bibinfo {pages} {072001} (\bibinfo {year} {2022}{\natexlab{a}})}\BibitemShut {NoStop}%
\bibitem [{\citenamefont {Frühwirth}\ \emph {et~al.}(2012)\citenamefont {Frühwirth}, \citenamefont {Neudecker},\ and\ \citenamefont {Leeb}}]{fruhwirth2012peelle}%
  \BibitemOpen
  \bibfield  {author} {\bibinfo {author} {\bibfnamefont {R.}~\bibnamefont {Frühwirth}}, \bibinfo {author} {\bibfnamefont {D.}~\bibnamefont {Neudecker}},\ and\ \bibinfo {author} {\bibfnamefont {H.}~\bibnamefont {Leeb}},\ }\bibfield  {title} {\bibinfo {title} {Peelle’s pertinent puzzle and its solution},\ }in\ \href {https://doi.org/10.1051/epjconf/20122700008} {\emph {\bibinfo {booktitle} {EPJ Web of Conferences}}},\ Vol.~\bibinfo {volume} {27}\ (\bibinfo  {publisher} {EDP Sciences},\ \bibinfo {year} {2012})\ p.\ \bibinfo {pages} {00008}\BibitemShut {NoStop}%
\bibitem [{\citenamefont {Mention}\ \emph {et~al.}(2011)\citenamefont {Mention}, \citenamefont {Fechner}, \citenamefont {Lasserre}, \citenamefont {Mueller}, \citenamefont {Lhuillier}, \citenamefont {Cribier},\ and\ \citenamefont {Letourneau}}]{Mention:2011rk}%
  \BibitemOpen
  \bibfield  {author} {\bibinfo {author} {\bibfnamefont {G.}~\bibnamefont {Mention}}, \bibinfo {author} {\bibfnamefont {M.}~\bibnamefont {Fechner}}, \bibinfo {author} {\bibfnamefont {T.}~\bibnamefont {Lasserre}}, \bibinfo {author} {\bibfnamefont {T.~A.}\ \bibnamefont {Mueller}}, \bibinfo {author} {\bibfnamefont {D.}~\bibnamefont {Lhuillier}}, \bibinfo {author} {\bibfnamefont {M.}~\bibnamefont {Cribier}},\ and\ \bibinfo {author} {\bibfnamefont {A.}~\bibnamefont {Letourneau}},\ }\bibfield  {title} {\bibinfo {title} {{The Reactor Antineutrino Anomaly}},\ }\href {https://doi.org/10.1103/PhysRevD.83.073006} {\bibfield  {journal} {\bibinfo  {journal} {Phys. Rev. D}\ }\textbf {\bibinfo {volume} {83}},\ \bibinfo {pages} {073006} (\bibinfo {year} {2011})}\BibitemShut {NoStop}%
\bibitem [{\citenamefont {Zhang}\ \emph {et~al.}(2024)\citenamefont {Zhang}, \citenamefont {Qian},\ and\ \citenamefont {Fallot}}]{Zhang:2023zif}%
  \BibitemOpen
  \bibfield  {author} {\bibinfo {author} {\bibfnamefont {C.}~\bibnamefont {Zhang}}, \bibinfo {author} {\bibfnamefont {X.}~\bibnamefont {Qian}},\ and\ \bibinfo {author} {\bibfnamefont {M.}~\bibnamefont {Fallot}},\ }\bibfield  {title} {\bibinfo {title} {{Reactor antineutrino flux and anomaly}},\ }\href {https://doi.org/10.1016/j.ppnp.2024.104106} {\bibfield  {journal} {\bibinfo  {journal} {Prog. Part. Nucl. Phys.}\ }\textbf {\bibinfo {volume} {136}},\ \bibinfo {pages} {104106} (\bibinfo {year} {2024})}\BibitemShut {NoStop}%
\bibitem [{\citenamefont {Zhang}\ \emph {et~al.}(2013)\citenamefont {Zhang}, \citenamefont {Qian},\ and\ \citenamefont {Vogel}}]{Zhang:2013ela}%
  \BibitemOpen
  \bibfield  {author} {\bibinfo {author} {\bibfnamefont {C.}~\bibnamefont {Zhang}}, \bibinfo {author} {\bibfnamefont {X.}~\bibnamefont {Qian}},\ and\ \bibinfo {author} {\bibfnamefont {P.}~\bibnamefont {Vogel}},\ }\bibfield  {title} {\bibinfo {title} {{Reactor Antineutrino Anomaly with known $\theta_{13}$}},\ }\href {https://doi.org/10.1103/PhysRevD.87.073018} {\bibfield  {journal} {\bibinfo  {journal} {Phys. Rev. D}\ }\textbf {\bibinfo {volume} {87}},\ \bibinfo {pages} {073018} (\bibinfo {year} {2013})}\BibitemShut {NoStop}%
\bibitem [{\citenamefont {Giunti}\ \emph {et~al.}(2022)\citenamefont {Giunti}, \citenamefont {Li}, \citenamefont {Ternes},\ and\ \citenamefont {Xin}}]{GIUNTI2022137054}%
  \BibitemOpen
  \bibfield  {author} {\bibinfo {author} {\bibfnamefont {C.}~\bibnamefont {Giunti}}, \bibinfo {author} {\bibfnamefont {Y.}~\bibnamefont {Li}}, \bibinfo {author} {\bibfnamefont {C.}~\bibnamefont {Ternes}},\ and\ \bibinfo {author} {\bibfnamefont {Z.}~\bibnamefont {Xin}},\ }\bibfield  {title} {\bibinfo {title} {Reactor antineutrino anomaly in light of recent flux model refinements},\ }\href {https://doi.org/https://doi.org/10.1016/j.physletb.2022.137054} {\bibfield  {journal} {\bibinfo  {journal} {Physics Letters B}\ }\textbf {\bibinfo {volume} {829}},\ \bibinfo {pages} {137054} (\bibinfo {year} {2022})}\BibitemShut {NoStop}%
\bibitem [{\citenamefont {Chakrani}\ \emph {et~al.}(2024)\citenamefont {Chakrani} \emph {et~al.}}]{Chakrani:2023htw}%
  \BibitemOpen
  \bibfield  {author} {\bibinfo {author} {\bibfnamefont {J.}~\bibnamefont {Chakrani}} \emph {et~al.},\ }\bibfield  {title} {\bibinfo {title} {{Parametrized uncertainties in the spectral function model of neutrino charged-current quasielastic interactions for oscillation analyses}},\ }\href {https://doi.org/10.1103/PhysRevD.109.072006} {\bibfield  {journal} {\bibinfo  {journal} {Phys. Rev. D}\ }\textbf {\bibinfo {volume} {109}},\ \bibinfo {pages} {072006} (\bibinfo {year} {2024})}\BibitemShut {NoStop}%
\bibitem [{\citenamefont {D'Agostini}(1994)}]{DAgostini:1993arp}%
  \BibitemOpen
  \bibfield  {author} {\bibinfo {author} {\bibfnamefont {G.}~\bibnamefont {D'Agostini}},\ }\bibfield  {title} {\bibinfo {title} {{On the use of the covariance matrix to fit correlated data}},\ }\href {https://doi.org/10.1016/0168-9002(94)90719-6} {\bibfield  {journal} {\bibinfo  {journal} {Nucl. Instrum. Meth. A}\ }\textbf {\bibinfo {volume} {346}},\ \bibinfo {pages} {306} (\bibinfo {year} {1994})}\BibitemShut {NoStop}%
\bibitem [{\citenamefont {Radev}\ and\ \citenamefont {Dolan}(2024)}]{radev2024flow}%
  \BibitemOpen
  \bibfield  {author} {\bibinfo {author} {\bibfnamefont {R.}~\bibnamefont {Radev}}\ and\ \bibinfo {author} {\bibfnamefont {S.}~\bibnamefont {Dolan}},\ }\href {https://agenda.infn.it/event/37867/contributions/227708/} {\bibinfo {title} {Flow matching mitigates gaussian error approximations in neutrino cross-section measurements}} (\bibinfo {year} {2024}),\ \bibinfo {note} {presented at International Workshop on Neutrino Cross Sections for Astrophysics and Oscillations (NuSTEC 2024)}\BibitemShut {NoStop}%
\bibitem [{\citenamefont {Abratenko}\ \emph {et~al.}(2025)\citenamefont {Abratenko} \emph {et~al.}}]{MicroBooNE:2024kwe}%
  \BibitemOpen
  \bibfield  {author} {\bibinfo {author} {\bibfnamefont {P.}~\bibnamefont {Abratenko}} \emph {et~al.} (\bibinfo {collaboration} {MicroBooNE}),\ }\bibfield  {title} {\bibinfo {title} {{Data-driven model validation for neutrino-nucleus cross section measurements}},\ }\href {https://doi.org/10.1103/PhysRevD.111.092010} {\bibfield  {journal} {\bibinfo  {journal} {Phys. Rev. D}\ }\textbf {\bibinfo {volume} {111}},\ \bibinfo {pages} {092010} (\bibinfo {year} {2025})}\BibitemShut {NoStop}%
\bibitem [{\citenamefont {Abratenko}\ \emph {et~al.}(2022{\natexlab{b}})\citenamefont {Abratenko} \emph {et~al.}}]{MicroBooNE:2021sfa}%
  \BibitemOpen
  \bibfield  {author} {\bibinfo {author} {\bibfnamefont {P.}~\bibnamefont {Abratenko}} \emph {et~al.} (\bibinfo {collaboration} {MicroBooNE}),\ }\bibfield  {title} {\bibinfo {title} {{First Measurement of Energy-Dependent Inclusive Muon Neutrino Charged-Current Cross Sections on Argon with the MicroBooNE Detector}},\ }\href {https://doi.org/10.1103/PhysRevLett.128.151801} {\bibfield  {journal} {\bibinfo  {journal} {Phys. Rev. Lett.}\ }\textbf {\bibinfo {volume} {128}},\ \bibinfo {pages} {151801} (\bibinfo {year} {2022}{\natexlab{b}})}\BibitemShut {NoStop}%
\bibitem [{\citenamefont {Abratenko}\ \emph {et~al.}(2023)\citenamefont {Abratenko} \emph {et~al.}}]{MicroBooNE:2023foc}%
  \BibitemOpen
  \bibfield  {author} {\bibinfo {author} {\bibfnamefont {P.}~\bibnamefont {Abratenko}} \emph {et~al.} (\bibinfo {collaboration} {MicroBooNE}),\ }\href@noop {} {\bibinfo {title} {{Measurement of three-dimensional inclusive muon-neutrino charged-current cross sections on argon with the MicroBooNE detector}}} (\bibinfo {year} {2023}),\ \Eprint {https://arxiv.org/abs/2307.06413} {arXiv:2307.06413 [hep-ex]} \BibitemShut {NoStop}%
\bibitem [{\citenamefont {Abratenko}\ \emph {et~al.}(2024{\natexlab{a}})\citenamefont {Abratenko} \emph {et~al.}}]{MicroBooNE:2024xod}%
  \BibitemOpen
  \bibfield  {author} {\bibinfo {author} {\bibfnamefont {P.}~\bibnamefont {Abratenko}} \emph {et~al.} (\bibinfo {collaboration} {MicroBooNE}),\ }\bibfield  {title} {\bibinfo {title} {{Inclusive cross section measurements in final states with and without protons for charged-current \ensuremath{\nu_\mu}-Ar scattering in MicroBooNE}},\ }\href {https://doi.org/10.1103/PhysRevD.110.013006} {\bibfield  {journal} {\bibinfo  {journal} {Phys. Rev. D}\ }\textbf {\bibinfo {volume} {110}},\ \bibinfo {pages} {013006} (\bibinfo {year} {2024}{\natexlab{a}})}\BibitemShut {NoStop}%
\bibitem [{\citenamefont {Tang}\ \emph {et~al.}(2017)\citenamefont {Tang}, \citenamefont {Li}, \citenamefont {Qian}, \citenamefont {Wei},\ and\ \citenamefont {Zhang}}]{Tang:2017rob}%
  \BibitemOpen
  \bibfield  {author} {\bibinfo {author} {\bibfnamefont {W.}~\bibnamefont {Tang}}, \bibinfo {author} {\bibfnamefont {X.}~\bibnamefont {Li}}, \bibinfo {author} {\bibfnamefont {X.}~\bibnamefont {Qian}}, \bibinfo {author} {\bibfnamefont {H.}~\bibnamefont {Wei}},\ and\ \bibinfo {author} {\bibfnamefont {C.}~\bibnamefont {Zhang}},\ }\bibfield  {title} {\bibinfo {title} {{Data Unfolding with Wiener-SVD Method}},\ }\href {https://doi.org/10.1088/1748-0221/12/10/P10002} {\bibfield  {journal} {\bibinfo  {journal} {JINST}\ }\textbf {\bibinfo {volume} {12}}\bibinfo  {number} { (10)},\ \bibinfo {pages} {P10002}}\BibitemShut {NoStop}%
\bibitem [{\citenamefont {Gardiner}(2024)}]{gardiner2024}%
  \BibitemOpen
\bibfield  {number} {  }\bibfield  {author} {\bibinfo {author} {\bibfnamefont {S.}~\bibnamefont {Gardiner}},\ }\href {https://arxiv.org/abs/2401.04065} {\bibinfo {title} {Mathematical methods for neutrino cross-section extraction}} (\bibinfo {year} {2024}),\ \Eprint {https://arxiv.org/abs/2401.04065} {arXiv:2401.04065 [hep-ex]} \BibitemShut {NoStop}%
\bibitem [{\citenamefont {Koch}\ and\ \citenamefont {Dolan}(2020)}]{Koch:2020oyf}%
  \BibitemOpen
  \bibfield  {author} {\bibinfo {author} {\bibfnamefont {L.}~\bibnamefont {Koch}}\ and\ \bibinfo {author} {\bibfnamefont {S.}~\bibnamefont {Dolan}},\ }\bibfield  {title} {\bibinfo {title} {{Treatment of flux shape uncertainties in unfolded, flux-averaged neutrino cross-section measurements}},\ }\href {https://doi.org/10.1103/PhysRevD.102.113012} {\bibfield  {journal} {\bibinfo  {journal} {Phys. Rev. D}\ }\textbf {\bibinfo {volume} {102}},\ \bibinfo {pages} {113012} (\bibinfo {year} {2020})}\BibitemShut {NoStop}%
\bibitem [{\citenamefont {Abe}\ \emph {et~al.}(2018{\natexlab{b}})\citenamefont {Abe} \emph {et~al.}}]{PhysRevD.98.032003}%
  \BibitemOpen
  \bibfield  {author} {\bibinfo {author} {\bibfnamefont {K.}~\bibnamefont {Abe}} \emph {et~al.} (\bibinfo {collaboration} {The T2K Collaboration}),\ }\bibfield  {title} {\bibinfo {title} {Characterization of nuclear effects in muon-neutrino scattering on hydrocarbon with a measurement of final-state kinematics and correlations in charged-current pionless interactions at t2k},\ }\href {https://doi.org/10.1103/PhysRevD.98.032003} {\bibfield  {journal} {\bibinfo  {journal} {Phys. Rev. D}\ }\textbf {\bibinfo {volume} {98}},\ \bibinfo {pages} {032003} (\bibinfo {year} {2018}{\natexlab{b}})}\BibitemShut {NoStop}%
\bibitem [{\citenamefont {Rasmussen}\ and\ \citenamefont {Williams}(2006)}]{gp_rasmussen}%
  \BibitemOpen
  \bibfield  {author} {\bibinfo {author} {\bibfnamefont {C.~E.}\ \bibnamefont {Rasmussen}}\ and\ \bibinfo {author} {\bibfnamefont {C.~K.~I.}\ \bibnamefont {Williams}},\ }\href@noop {} {\emph {\bibinfo {title} {{Gaussian Processes for Machine Learning}}}}\ (\bibinfo  {publisher} {The MIT Press},\ \bibinfo {year} {2006})\BibitemShut {NoStop}%
\bibitem [{\citenamefont {James}\ and\ \citenamefont {Winkler}(2004)}]{James:2004xla}%
  \BibitemOpen
  \bibfield  {author} {\bibinfo {author} {\bibfnamefont {F.}~\bibnamefont {James}}\ and\ \bibinfo {author} {\bibfnamefont {M.}~\bibnamefont {Winkler}},\ }\href@noop {} {\bibinfo {title} {{MINUIT User's Guide}}} (\bibinfo {year} {2004})\BibitemShut {NoStop}%
\bibitem [{\citenamefont {Cooper-Troendle}(2025)}]{conditional_constraint_example}%
  \BibitemOpen
  \bibfield  {author} {\bibinfo {author} {\bibfnamefont {L.}~\bibnamefont {Cooper-Troendle}},\ }\href {https://github.com/London2/Conditional_Constraint_Fitting} {\bibinfo {title} {Conditional constraint fitting}} (\bibinfo {year} {2025})\BibitemShut {NoStop}%
\bibitem [{\citenamefont {Gross}\ and\ \citenamefont {Vitells}(2010)}]{Gross:2010qma}%
  \BibitemOpen
  \bibfield  {author} {\bibinfo {author} {\bibfnamefont {E.}~\bibnamefont {Gross}}\ and\ \bibinfo {author} {\bibfnamefont {O.}~\bibnamefont {Vitells}},\ }\bibfield  {title} {\bibinfo {title} {{Trial factors for the look elsewhere effect in high energy physics}},\ }\href {https://doi.org/10.1140/epjc/s10052-010-1470-8} {\bibfield  {journal} {\bibinfo  {journal} {Eur. Phys. J. C}\ }\textbf {\bibinfo {volume} {70}},\ \bibinfo {pages} {525} (\bibinfo {year} {2010})}\BibitemShut {NoStop}%
\bibitem [{\citenamefont {Sidák}(1967)}]{Sidak:1967}%
  \BibitemOpen
  \bibfield  {author} {\bibinfo {author} {\bibfnamefont {Z.}~\bibnamefont {Sidák}},\ }\bibfield  {title} {\bibinfo {title} {Rectangular confidence regions for the means of multivariate normal distributions},\ }\href {https://doi.org/10.2307/2283989} {\bibfield  {journal} {\bibinfo  {journal} {Journal of the American Statistical Association}\ }\textbf {\bibinfo {volume} {62}},\ \bibinfo {pages} {626} (\bibinfo {year} {1967})}\BibitemShut {NoStop}%
\bibitem [{\citenamefont {David}\ and\ \citenamefont {Johnson}(1948)}]{quantile_mapping}%
  \BibitemOpen
  \bibfield  {author} {\bibinfo {author} {\bibfnamefont {F.~N.}\ \bibnamefont {David}}\ and\ \bibinfo {author} {\bibfnamefont {N.~L.}\ \bibnamefont {Johnson}},\ }\bibfield  {title} {\bibinfo {title} {{The Probability Integral Transformation When Parameters are Estimated from the Sample}},\ }\href {https://doi.org/10.2307/2332638} {\bibfield  {journal} {\bibinfo  {journal} {Biometrika}\ }\textbf {\bibinfo {volume} {35}},\ \bibinfo {pages} {182} (\bibinfo {year} {1948})}\BibitemShut {NoStop}%
\bibitem [{\citenamefont {D'Agostini}(1995)}]{DAgostini:1994fjx}%
  \BibitemOpen
  \bibfield  {author} {\bibinfo {author} {\bibfnamefont {G.}~\bibnamefont {D'Agostini}},\ }\bibfield  {title} {\bibinfo {title} {{A Multidimensional unfolding method based on Bayes' theorem}},\ }\href {https://doi.org/10.1016/0168-9002(95)00274-X} {\bibfield  {journal} {\bibinfo  {journal} {Nucl. Instrum. Meth. A}\ }\textbf {\bibinfo {volume} {362}},\ \bibinfo {pages} {487} (\bibinfo {year} {1995})}\BibitemShut {NoStop}%
\bibitem [{\citenamefont {Abratenko}\ \emph {et~al.}(2024{\natexlab{b}})\citenamefont {Abratenko} \emph {et~al.}}]{PhysRevLett.133.041801}%
  \BibitemOpen
  \bibfield  {author} {\bibinfo {author} {\bibfnamefont {P.}~\bibnamefont {Abratenko}} \emph {et~al.} (\bibinfo {collaboration} {MicroBooNE Collaboration}),\ }\bibfield  {title} {\bibinfo {title} {First simultaneous measurement of differential muon-neutrino charged-current cross sections on argon for final states with and without protons using microboone data},\ }\href {https://doi.org/10.1103/PhysRevLett.133.041801} {\bibfield  {journal} {\bibinfo  {journal} {Phys. Rev. Lett.}\ }\textbf {\bibinfo {volume} {133}},\ \bibinfo {pages} {041801} (\bibinfo {year} {2024}{\natexlab{b}})}\BibitemShut {NoStop}%
\bibitem [{\citenamefont {Stowell}\ \emph {et~al.}(2017)\citenamefont {Stowell} \emph {et~al.}}]{Stowell:2016jfr}%
  \BibitemOpen
  \bibfield  {author} {\bibinfo {author} {\bibfnamefont {P.}~\bibnamefont {Stowell}} \emph {et~al.},\ }\bibfield  {title} {\bibinfo {title} {{NUISANCE: a neutrino cross-section generator tuning and comparison framework}},\ }\href {https://doi.org/10.1088/1748-0221/12/01/P01016} {\bibfield  {journal} {\bibinfo  {journal} {JINST}\ }\textbf {\bibinfo {volume} {12}},\ \bibinfo {pages} {P01016}}\BibitemShut {NoStop}%
\bibitem [{\citenamefont {{T2K Collaboration}}(2016)}]{T2K:2016jor}%
  \BibitemOpen
  \bibfield  {author} {\bibinfo {author} {\bibnamefont {{T2K Collaboration}}} (\bibinfo {collaboration} {T2K Collaboration}),\ }\bibfield  {title} {\bibinfo {title} {{Measurement of double-differential muon neutrino charged-current interactions on C$_8$H$_8$ without pions in the final state using the T2K off-axis beam}},\ }\href {https://doi.org/10.1103/PhysRevD.93.112012} {\bibfield  {journal} {\bibinfo  {journal} {Phys. Rev. D}\ }\textbf {\bibinfo {volume} {93}},\ \bibinfo {pages} {112012} (\bibinfo {year} {2016})}\BibitemShut {NoStop}%
\bibitem [{\citenamefont {Abe}\ \emph {et~al.}(2020{\natexlab{b}})\citenamefont {Abe} \emph {et~al.}}]{T2K:2020jav}%
  \BibitemOpen
  \bibfield  {author} {\bibinfo {author} {\bibfnamefont {K.}~\bibnamefont {Abe}} \emph {et~al.} (\bibinfo {collaboration} {T2K}),\ }\bibfield  {title} {\bibinfo {title} {{Simultaneous measurement of the muon neutrino charged-current cross section on oxygen and carbon without pions in the final state at T2K}},\ }\href {https://doi.org/10.1103/PhysRevD.101.112004} {\bibfield  {journal} {\bibinfo  {journal} {Phys. Rev. D}\ }\textbf {\bibinfo {volume} {101}},\ \bibinfo {pages} {112004} (\bibinfo {year} {2020}{\natexlab{b}})}\BibitemShut {NoStop}%
\bibitem [{\citenamefont {Benhar}\ \emph {et~al.}(1994)\citenamefont {Benhar}, \citenamefont {Fabrocini}, \citenamefont {Fantoni},\ and\ \citenamefont {Sick}}]{Benhar:1994hw}%
  \BibitemOpen
  \bibfield  {author} {\bibinfo {author} {\bibfnamefont {O.}~\bibnamefont {Benhar}}, \bibinfo {author} {\bibfnamefont {A.}~\bibnamefont {Fabrocini}}, \bibinfo {author} {\bibfnamefont {S.}~\bibnamefont {Fantoni}},\ and\ \bibinfo {author} {\bibfnamefont {I.}~\bibnamefont {Sick}},\ }\bibfield  {title} {\bibinfo {title} {{Spectral function of finite nuclei and scattering of GeV electrons}},\ }\href {https://doi.org/10.1016/0375-9474(94)90920-2} {\bibfield  {journal} {\bibinfo  {journal} {Nucl. Phys.}\ }\textbf {\bibinfo {volume} {A579}},\ \bibinfo {pages} {493} (\bibinfo {year} {1994})}\BibitemShut {NoStop}%
\bibitem [{\citenamefont {Ankowski}\ \emph {et~al.}(2015)\citenamefont {Ankowski}, \citenamefont {Benhar},\ and\ \citenamefont {Sakuda}}]{Ankowski:2014yfa}%
  \BibitemOpen
  \bibfield  {author} {\bibinfo {author} {\bibfnamefont {A.~M.}\ \bibnamefont {Ankowski}}, \bibinfo {author} {\bibfnamefont {O.}~\bibnamefont {Benhar}},\ and\ \bibinfo {author} {\bibfnamefont {M.}~\bibnamefont {Sakuda}},\ }\bibfield  {title} {\bibinfo {title} {{Improving the accuracy of neutrino energy reconstruction in charged-current quasielastic scattering off nuclear targets}},\ }\href {https://doi.org/10.1103/PhysRevD.91.033005} {\bibfield  {journal} {\bibinfo  {journal} {Phys. Rev.}\ }\textbf {\bibinfo {volume} {D91}},\ \bibinfo {pages} {033005} (\bibinfo {year} {2015})}\BibitemShut {NoStop}%
\bibitem [{\citenamefont {Ankowski}\ and\ \citenamefont {Sobczyk}(2008)}]{Ankowski:2007uy}%
  \BibitemOpen
  \bibfield  {author} {\bibinfo {author} {\bibfnamefont {A.~M.}\ \bibnamefont {Ankowski}}\ and\ \bibinfo {author} {\bibfnamefont {J.~T.}\ \bibnamefont {Sobczyk}},\ }\bibfield  {title} {\bibinfo {title} {{Construction of spectral functions for medium-mass nuclei}},\ }\href {https://doi.org/10.1103/PhysRevC.77.044311} {\bibfield  {journal} {\bibinfo  {journal} {Phys. Rev. C}\ }\textbf {\bibinfo {volume} {77}},\ \bibinfo {pages} {044311} (\bibinfo {year} {2008})}\BibitemShut {NoStop}%
\bibitem [{\citenamefont {Graczyk}\ and\ \citenamefont {Sobczyk}(2008)}]{SobczykGraczyk}%
  \BibitemOpen
  \bibfield  {author} {\bibinfo {author} {\bibfnamefont {K.~M.}\ \bibnamefont {Graczyk}}\ and\ \bibinfo {author} {\bibfnamefont {J.~T.}\ \bibnamefont {Sobczyk}},\ }\bibfield  {title} {\bibinfo {title} {{Form Factors in the Quark Resonance Model}},\ }\href {https://doi.org/10.1103/PhysRevD.79.079903} {\bibfield  {journal} {\bibinfo  {journal} {Phys. Rev. D}\ }\textbf {\bibinfo {volume} {77}},\ \bibinfo {pages} {053001} (\bibinfo {year} {2008})},\ \bibinfo {note} {[Erratum: Phys.Rev.D 79, 079903 (2009)]}\BibitemShut {NoStop}%
\bibitem [{\citenamefont {Nieves}\ \emph {et~al.}(2011)\citenamefont {Nieves}, \citenamefont {Ruiz~Simo},\ and\ \citenamefont {Vicente~Vacas}}]{Nieves:2011pp}%
  \BibitemOpen
  \bibfield  {author} {\bibinfo {author} {\bibfnamefont {J.}~\bibnamefont {Nieves}}, \bibinfo {author} {\bibfnamefont {I.}~\bibnamefont {Ruiz~Simo}},\ and\ \bibinfo {author} {\bibfnamefont {M.~J.}\ \bibnamefont {Vicente~Vacas}},\ }\bibfield  {title} {\bibinfo {title} {{Inclusive Charged--Current Neutrino--Nucleus Reactions}},\ }\href {https://doi.org/10.1103/PhysRevC.83.045501} {\bibfield  {journal} {\bibinfo  {journal} {Phys. Rev. C}\ }\textbf {\bibinfo {volume} {83}},\ \bibinfo {pages} {045501} (\bibinfo {year} {2011})}\BibitemShut {NoStop}%
\bibitem [{\citenamefont {Abe}\ \emph {et~al.}(2013)\citenamefont {Abe} \emph {et~al.}}]{PhysRevD.87.012001}%
  \BibitemOpen
  \bibfield  {author} {\bibinfo {author} {\bibfnamefont {K.}~\bibnamefont {Abe}} \emph {et~al.} (\bibinfo {collaboration} {T2K Collaboration}),\ }\bibfield  {title} {\bibinfo {title} {T2k neutrino flux prediction},\ }\href {https://doi.org/10.1103/PhysRevD.87.012001} {\bibfield  {journal} {\bibinfo  {journal} {Phys. Rev. D}\ }\textbf {\bibinfo {volume} {87}},\ \bibinfo {pages} {012001} (\bibinfo {year} {2013})}\BibitemShut {NoStop}%
\end{thebibliography}
%

\end{document}